\documentclass[sigconf]{acmart}
\usepackage[normalem]{ulem}
\useunder{\uline}{\ul}{}
\usepackage{graphicx}
\usepackage{subcaption}
\usepackage{amsmath}
\usepackage{enumitem}
\usepackage{xspace}
\usepackage{balance}
\usepackage{pifont} % for \ding symbols
\usepackage{array} 
\usepackage[most]{tcolorbox}
\usepackage{diagbox} 
\usepackage{tabularx}
\usepackage{lipsum} % for dummy text

\newtcolorbox{promptbox}{
  colback=gray!15,    % background
  colframe=gray!15,   % same as background (no visible border)
  boxrule=0pt,        % remove border line
  left=0mm, right=0mm, top=0mm, bottom=0mm
}

\theoremstyle{definition}

\newcommand{\modelname}{AgentDR\xspace}
\newcommand{\cmark}{\ding{51}}% ✓
\newcommand{\xmark}{\ding{55}}% ✗
\AtBeginDocument{%
  }

\copyrightyear{2026}
\acmYear{2026}
\setcopyright{cc}
\setcctype{by}
\acmConference[WWW '26] {Proceedings of the ACM Web Conference 2026}{April 13--17, 2026}{Dubai, United Arab Emirates.}
\acmBooktitle{Proceedings of the ACM Web Conference 2026 (WWW '26), April 13--17, 2026, Dubai, United Arab Emirates}
\acmISBN{979-8-4007-2307-0/2026/04}
\acmDOI{10.1145/3774904.3792304}
\begin{document}

% \title{Cross-view Graph Self-supervised Learning for Group Identification via Transitional Hypergraph Convolution}
% \title{Unified Multitask pretraining for Recommendation via Hypergraph with Transitional Attention}
\title{AgentDR: Dynamic Recommendation with Implicit Item-Item Relations via LLM-based Agents}

\settopmatter{authorsperrow=4}

\author{Mingdai~Yang}
\email{myang72@uic.edu}
\affiliation{%
  \institution{Univ.\ of Illinois  Chicago}
  \city{Chicago}
  \state{IL}
  \country{USA}}
  
\author{Nurendra~Choudhary}
\email{nurendc@amazon.com}
\affiliation{%
  \institution{Amazon}
  \city{Mountain View}
  \state{CA}
  \country{USA}}
  
\author{Jiangshu~Du}
\email{jiangshd@amazon.com}
\affiliation{%
  \institution{Amazon}
  \city{Mountain View}
  \state{CA}
  \country{USA}}

\author{Edward W~Huang}
\email{ewhuang@amazon.com}
\affiliation{%
  \institution{Amazon}
  \city{Mountain View}
  \state{CA}
  \country{USA}}
  
\author{Philip Yu}
\email{psyu@uic.edu}
\affiliation{%
  \institution{Univ.\ of Illinois  Chicago}
  \city{Chicago}
  \state{IL}
  \country{USA}}
  
\author{Karthik~Subbian}
\email{ksubbian@amazon.com}
\affiliation{%
  \institution{Amazon}
  \city{Mountain View}
  \state{CA}
  \country{USA}} 
  
\author{Danai~Koutra}
\email{dkoutra@umich.edu}
\affiliation{%
  \institution{University of Michigan}
  \city{Ann Arbor}
  \state{MI}
  \country{USA}} 
\authornote{Danai Koutra holds concurrent appointments as an Amazon Scholar and a Professor at UM. This paper describes work performed at Amazon.}
  
\renewcommand{\shortauthors}{Mingdai Yang et al.}



\begin{abstract}
 % Traditional recommender systems often overlook the rich semantic information in textual content, while recent efforts to incorporate large language models (LLMs) as either feature encoders or standalone recommenders face limitations in representation learning and behavior pattern capturing. 
 Recent agent-based recommendation frameworks aim to simulate user behaviors by incorporating memory mechanisms and prompting strategies, but they struggle with hallucinating non-existent items and full-catalog ranking.
 Besides, a largely underexplored opportunity lies in leveraging LLMs’ commonsense reasoning to capture user intent through substitute and complement relationships between items, which are usually implicit in datasets and difficult for traditional ID-based recommenders to capture.
 In this work, we propose a novel LLM-agent framework, \textbf{\modelname}, which bridges LLM reasoning with scalable recommendation tools. Our approach delegates full-ranking tasks to traditional models while utilizing LLMs to (i) integrate multiple recommendation outputs based on personalized tool suitability and (ii) reason over substitute and complement relationships grounded in user history. This design mitigates hallucination, scales to large catalogs, and enhances recommendation relevance through relational reasoning. Through extensive experiments on three public grocery datasets, we show that our framework achieves superior full-ranking performance, yielding on average a twofold improvement over its underlying tools. We also introduce a new LLM-based evaluation metric that jointly measures semantic alignment and ranking correctness.
\end{abstract}

\begin{CCSXML}
<ccs2012>
   <concept>
       <concept_id>10002951.10003317.10003338</concept_id>
       <concept_desc>Information systems~Retrieval models and ranking</concept_desc>
       <concept_significance>500</concept_significance>
       </concept>
 </ccs2012>
\end{CCSXML}

\ccsdesc[500]{Information systems~Retrieval models and ranking}

\keywords{Agents; Recommender Systems; Large Language Models}


\maketitle

% Some Space Left in advan sce for Names of All Authors
\section{Introduction}
Recommender systems provide personalized suggestions to assist users in information discovery and decision-making~\cite{cfag,uprth,gtgs,Loveland25-CF}. However, traditional ID-based recommender systems~\cite{LightGCN, sasrec, enmf} solely based on user-item interactions are constrained by overlooking the rich semantic information contained in textual context. Even when LLMs are employed to encode textual features into embeddings~\cite{llmrec, rlmrec,Zhu_2025_CVPR}, these fixed-length representations inherently struggle to preserve the full semantic richness of textual information for recommendation~\cite{0007XCW25, NEURIPS2024_2db8ce96}. There are several attempts to harness the commonsense reasoning and knowledge utilization capabilities of LLMs by directly applying them as recommender systems, reformulating recommendation tasks as language modeling problems~\cite{p5, llmrank, m6rec,Zhongmou25-Linkgpt}. However, an inherent gap remains because LLMs are optimized for next-token prediction in natural language rather than learning collaborative and sequential patterns from interactions, limiting their efficiency in user behavior modeling.

\begin{figure}[]
    \centering
    \includegraphics[scale=0.35]{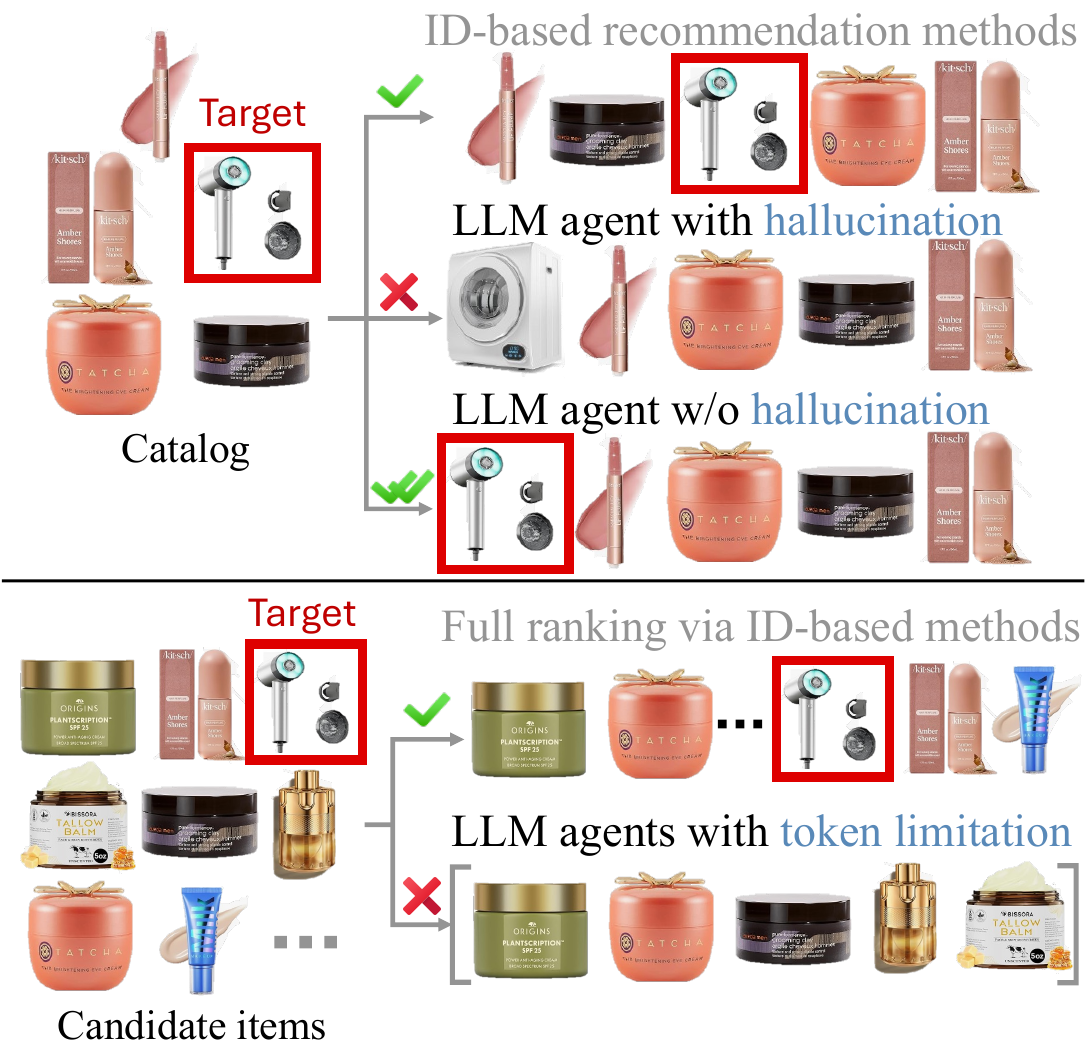}
    % \caption{Two major drawbacks in directly deploying LLMs for recommendation: LLMs can hallucinate items not existing in the actual product catalog, and token limits of LLMs make them unsuitable for ranking items at scale. \modelname addresses both limitations by delegating full-ranking tasks to recommendation tools.}
    \caption{There are two major drawbacks in directly deploying LLMs for recommendation: hallucination and token limitation. \modelname addresses them by delegating full-ranking tasks to recommendation tools.}
    % \caption{Two major drawbacks in directly deploying LLMs for recommendation: hallucination and token limits.}
    \label{fig:motiv}
    \vspace{-3mm}
\end{figure}

Given the limitations of using LLMs solely as feature encoders or standalone recommenders, recent studies have investigated leveraging LLMs as interactive agents for recommendation~\cite{agentcf, agent4rec, recagent}. Rather than relying only on direct prompting with user interaction histories, these agent-based approaches integrate memory modules to preserve long-term context and better harness collaborative signals throughout the recommendation process. 
% During training, the agent memories are dynamically updated based on observed behavioral patterns through a reflection mechanism, allowing the agent to iteratively refine its internal representation of user preferences. 
While LLM-based agents alleviate the representational constraints of static numerical embeddings and enable adaptive modeling of user behavior over time, they also inherit key limitations from the generative nature of LLMs. As illustrated in Fig.~\ref{fig:motiv}, two primary drawbacks arise in previous agent-based works when directly applying LLMs for recommendation: (i) the tendency to hallucinate non-existent items that are not part of the product catalog, and (ii) the high inference cost and token length constraints, which make them impractical for large-scale item ranking. Moreover, longer input and output sequences further increase the risk of hallucination due to attention bottlenecks and heightened generation uncertainty~\cite{hallulength1}.

As a result, when applied to recommendation scenarios, existing LLM-based agents~\cite{agentcf,iagent, agent4rec} typically evaluate performance by checking whether the ground-truth item, which is the next item that the user actually interacted, is correctly ranked against a small set of negative candidates. These negative candidates are items randomly sampled from the catalog that the user did not interact with during the observation window, serving as distractors in the ranking process. However, this setup is fundamentally misaligned with most real-world recommendation scenarios, where systems must select the most relevant items from massive catalogs that often contain thousands of items. Restricting the candidate set to such a small range can artificially inflate performance, overlook the complexity of large-scale ranking, and hinder the system’s ability to identify truly relevant items in open-ended domains.
% \begin{figure}[]
%     \centering
%     \includegraphics[scale=0.28]{figures/motivation/motivation2.pdf}
%     \caption{A toy example on recommendation enhancement via incorporating item-item relations through LLM agents.}
%     \label{fig:motiv2}
% \end{figure}

Besides their limitations in full-ranking, previous LLM-based agents primarily focus on modeling user preferences from explicit user-item interactions, which already can be effectively modeled by traditional embedding-based recommendation methods~\cite{simplex,enmf,LightGCN}. On the contrary, traditional models often overlook higher-level semantic relations between items that go beyond co-occurrence patterns. 
To address this gap, the generative capabilities and extensive world knowledge of LLMs make them particularly well-suited for reasoning over implicit relationships, such as complementarity and substitution relations.
Substitutes help identify alternative options when a preferred item is unavailable or when the user seeks variety, while complements reveal co-purchase patterns that often indicate planned or contextual consumption behaviors. 
For example, a user purchasing a digital single-lens reflex camera often requires a tripod as a complement, whereas a user who buys tiramisu is more likely to choose cheesecake as substitute over burgers when the original item is out of stock.
Although such relationships can greatly enhance recommendation quality~\cite{subcom22, subcom23,Nerrise25-substitution}, the ground-truth labels of them are typically absent from most datasets~\cite{lacklabel19, lacklabel21}. Furthermore, annotating item pairs with accurate relational labels often requires extensive world knowledge and nuanced understanding, making manual annotation both expensive and time-consuming. This gap highlights the potential of LLMs as an alternative source of relational knowledge. By leveraging their extensive pretraining on diverse textual data, LLMs can capture subtle functional relationships between items, enabling automatic discovery of substitutes and complements that would otherwise require labor-intensive human annotation.

Considering the two issues discussed above, we propose a novel LLM-based agent framework that leverages the complementary strengths of LLMs and traditional recommendation models. This framework addresses hallucination by grounding all recommendations in catalog-aware models. The reasoning ability and world knowledge of LLMs are utilized to integrate the ranking results of multiple recommendation tools and to perform fine-grained refinement of the combined ranking results, both in a personalized manner. For each user, an agent is optimized to assess the suitability of each recommendation tool by measuring the gap between the tool’s outputs and the user’s true preferences. In parallel, the user’s intent to purchase substitutes or complements is inferred by the LLM based on the user’s historical behavior. During inference, the agent first leverages the stored tool suitability to integrate the outputs from all tools, and then applies the inferred user intent to refine the combined ranking results.
Building on this framework, we highlight the main contributions of our work as follows:
\begin{itemize}[left=0pt]
    \item \textbf{Novel Framework.} We propose an agent-based framework \modelname, where LLM-based agents delegate full-ranking recommendation to recommendation tools that excel at modeling complex user behavior patterns. This separation enables the integration of the scalability and efficiency of traditional recommenders with the reasoning capabilities of LLMs.
    \item \textbf{Item-item Reasoning.} We leverage the world knowledge of LLMs for implicit item-item relationship reasoning to enhance recommendation. Each user agent generates substitute and complement candidates based on the user's historical preferences, and accordingly refines recommendation results.
    \item \textbf{Comprehensive Evaluation.} Extensive experiments verify the effectiveness of our approach in full-ranking performance on three public datasets. Besides recall and NDCG, we propose a LLM-based metric to jointly assess the semantic relevance and ordering correctness of the predicted ranking lists.
\end{itemize}

\section{Related Work}

\begin{table}
\caption{Comparison of recommendation approaches.}
\label{tab:salesman}
\vspace{-0.35cm}
\centering
\resizebox{0.44\textwidth}{!}{%
\begin{tabular}{l c c c}
\toprule
%Method
& Full Ranking
& Text Feature
& Generative  \\
& Evaluation
& Handling
& Reasoning\\
\midrule
Recommender based on interactions
& \cmark & \xmark & \xmark \\
LLM as Encoder
& \cmark & \cmark & \xmark \\
LLM as Recommender
& \xmark & \cmark & \cmark \\
LLM-Rec Agents~\cite{interecagent, recmind, agentcf, agent4rec, recagent, iagent}
& \xmark & \cmark & \cmark \\
\midrule
\modelname
& \cmark & \cmark & \cmark \\
\bottomrule
\end{tabular}%
}
\vspace{-0.4cm}
\end{table}

% previous version
% \begin{table}[]\caption{Comparison of recommendation approaches.}\label{tab:salesman}
% \vspace{-2mm}
% \centering
% \resizebox{0.48\textwidth}{!}{%
% \begin{tabular}{|l|m{0.1\textwidth} |m{0.09\textwidth} |m{0.09\textwidth}|}
% \hline
% \diagbox[width=19em,trim=l]{Method}{Capability} &
% \centering Full Ranking\\ Evaluation &
% \centering Text Feature\\ Handling &
% \centering Generative\\ Reasoning \tabularnewline
% \hline
% Recommender based on Interactions &\centering\cmark &\centering\xmark & \centering\arraybackslash\xmark  \\
% LLM as Encoder       &\centering\cmark &\centering\cmark  &\centering\arraybackslash\xmark  \\
% LLM as Recommender   &\centering\xmark &\centering\cmark  &\centering\arraybackslash\cmark\\
% LLM-Rec Agent~\cite{interecagent, recmind, agentcf, agent4rec, recagent, iagent} &\centering\xmark &\centering\cmark  &\centering\arraybackslash\cmark \\
% \hline
% \modelname            &\centering\cmark &\centering\cmark &\centering\arraybackslash\cmark \\
% \hline
% \end{tabular}}
% \vspace{-2mm}
% \end{table}

LLM-based agents have recently demonstrated strong capabilities in using memory and reflection to learn from past experiences and simulate human-like behaviors~\cite{gao2025,ZhongGGYW24,ShinnCGNY23, ZhaoTZC024, ShiJQY24, LiDWCF24}. In the context of recommendation, LLM-based agents have been used to investigate social phenomena such as information cocoons and filter bubble effect by simulating user behaviors~\cite{recagent,agent4rec}.
%Agent4Rec~\cite{agent4rec} reproduces the filter bubble effect and uncovers its underlying causal mechanisms. 
Given that agent-based frameworks can offer deeper insights into user behavior by leveraging the reasoning capabilities of LLMs, there is a growing interest in applying LLM-based agents to enhance recommendation performance~\cite{recmind, interecagent,agentcf, iagent}. 
% RecMind~\cite{recmind} leverages historical data to build user agents for few-shot recommendation. InteRecAgent~\cite{interecagent} connects LLM capabilities with tools for conversational recommendation. AgentCF~\cite{agentcf} employs collaborative reflection to update memory of both user and item agents. User instructions and reviews are also leveraged to flexibly express user preferences beyond just product attributes in iAgent~\cite{iagent}.
Nonetheless, since hallucination~\cite{hallu1, hallu2} and token limitation remain two challenges when applying these agents to rank all items in the datasets for each user, most prior works~\cite{agentcf,recmind,interecagent,agent4rec,recagent, iagent} lack of full-ranking ability. The batch-ranking task in these works is fundamentally misaligned with most real-world scenarios, where recommender systems are required to rank relevant items from massive catalogs. And these works focus on predicting user behaviors based on user-item interactions, a signal that traditional recommendation models~\cite{LightGCN, sasrec, simplex} are already well-equipped to capture and exploit. Different from these works, our framework combines the strengths of LLMs and traditional recommenders to achieve personalized full-ranking. Related works other than LLM-based agent frameworks for recommendation are discussed in Appendix ~\ref{app:related_works}. Comparison between different approaches is summarized in Table~\ref{tab:salesman}.

\section{Proposed Framework: \modelname}

\begin{figure*}[h]
    \centering
    \includegraphics[scale=0.4]{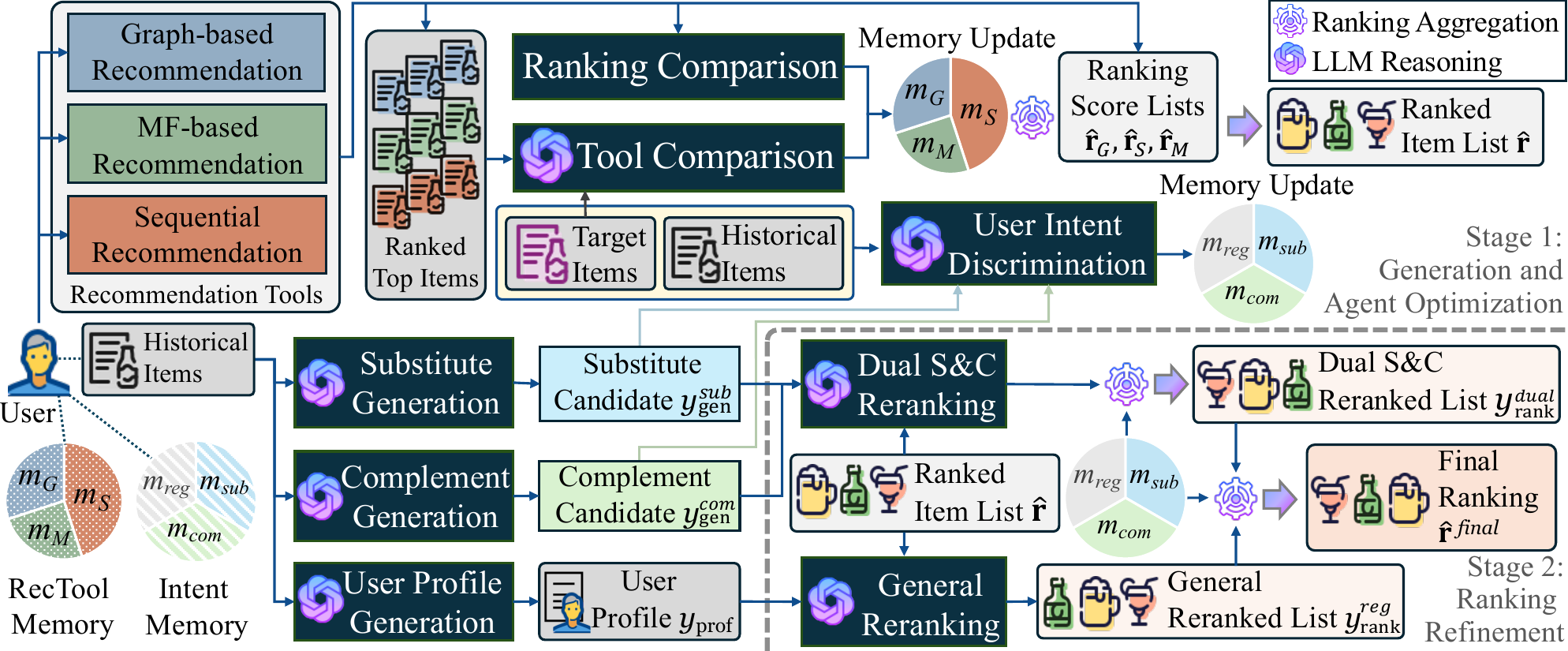}
    \caption{The framework of \modelname: Each user agent is equipped with two memory modules: RecTool memory to store tool suitability, and intent memory to track user intent. Ranking results from recommendation tools are aggregated based on RecTool memory. The aggregated result is further refined by the dual S\&C and general ranking modules.}
    \label{fig:framework}
\end{figure*}

\subsection{Preliminaries}
\subsubsection{Problem Formulation}
Let $\mathcal{I}$ and $\mathcal{U}$ denote the complete item pool and the set of users in the dataset, respectively. Specifically, each user $u \in \mathcal{U}$ is associated with a behavior sequence $\mathbf{s} = (i_1, i_2, \dots, i_n)$, where $i_j \in \mathcal{I}$ denotes the $j$-th item interacted with by user $u$. Furthermore, each item $i \in \mathcal{I}$ is accompanied by its corresponding textual description. For simplicity, we use $\text{desc}(\mathbf{s})$ to denote the list of textual descriptions for all items from item sequences $\mathbf{s}$.
Given a user's historical interaction sequence $\mathbf{s}_{[:-k]}$ and the corresponding item description list, the goal is to predict the next item $i_{k+1} \in \mathcal{I}$ that the user is likely to interact with. 
This is commonly framed as a ranking task: each user agent is expected to assign higher scores to items that better match the user's preferences, as inferred from both behavioral patterns and textual semantics. The final output is a ranked list of $\mathcal{I}$ tailored to this user $u$. For clarity, the subscript $u$ will be omitted in subsequent sections, given that each user agent functions independently.

\subsubsection{Recommendation Tools}~\label{sec:rectools}
AgentDR is an agent-based framework compatible with different recommendation methods as tools under the minimal assumption that they produce a full-ranking list of items. These recommendation methods are treated as tools and collectively form the tool set $\mathcal{T}$. For a broad coverage of recommendation modeling strategies, we utilize three types of recommendation tools to perform recommendation tasks within \modelname: (i) a graph-based model with output ranking score list $\hat{\mathbf{r}}_{G}$; (ii) a sequential recommendation model with output ranking score list $\hat{\mathbf{r}}_{S}\in\mathbb{R}$; and (iii) a MF-based model with output ranking score list $\hat{\mathbf{r}}_{M}$. All these output lists are in $\mathbb{R}^{1\times|\mathcal{I}|}$. The three model instances used in this work are elaborated in Sec.~\ref{sec:toolselect}. These models are pretrained on the first $(n-k)$ items, denoted as $\mathbf{s}_{[:-k]}$.

\subsubsection{Framework Overview} The overall framework of \modelname is depicted in Fig.~\ref{fig:framework}. Each user agent is equipped with a profile and two memory modules (Sec.~\ref{sec:profile_mem}). 
% The agent memory is optimized in the first stage and then used for ranking refinement in the second stage. 
The agent optimization begins with prompting the LLM to generate potential substitutes and complements for each user (Sec.~\ref{sec:scgen}). Next, the agent memory is optimized by distinct reflection mechanisms. The RecTool memory is optimized through personalized tool selection (Sec.~\ref{sec_tool_select}), while the intent memory is optimized via user intent discrimination (Sec.~\ref{sec:intent_compare}). Finally, the top items from the aggregated rankings of multiple tools are refined by dual S\&C reranking module (Sec.~\ref{sec:dualrerank}).

\subsection{User Profile and Memory}\label{sec:profile_mem}
\subsubsection{User Profile Summarization}~\label{sec:userprof}
Explicit user profiles are usually unavailable in most public real-world datasets. Thus, we summarize the user profile from users' historical interactions by leveraging the reasoning capabilities and world knowledge of LLMs. To be concrete, we prompt the LLM to summarize the user's preference from descriptions of historical items for each user. And this preference is used as the user profile $p$ of each user agent:
\begin{equation}\label{eq:gen_prof}
        y_{\text{prof}} = \text{LLM}_\text{prof}(\text{desc}(\mathbf{s})),
\end{equation}
where $\text{LLM}_\text{prof}(\cdot)$ represents the LLM prompted to transform these texts into a coherent user profile. Due to limited space, the details of all prompts used in \modelname are shown in Appendix~\ref{app:prompts}.

\subsubsection{User Agent Memory}
In contrast to static user profiles, memories are employed to enable user agents to dynamically maintain context in response to agent optimization based on consecutive user behaviors. 
% The relationship between user profiles and agent memory is analogous to the distinction between fixed one-hot feature vectors and trainable embeddings in traditional models. 
We equip each user agent with two compact memory modules: \textit{RecTool} memory and \textit{intent} memory. RecTool memory $\mathbf{m}^{rec}\in\mathbb{R}^{1\times |\mathcal{T}|}$ is a list of weights reflecting the suitability of each recommendation tool for each user, where $d$ is the number of recommendation tools. For LightGCN, SASRec and SimpleX deployed in this paper, we denote their weights in memory as $m_G$, $m_S$ and $m_M$, respectively. Similarly, intent memory $\mathbf{m}^{int}\in\mathbb{R}^{1\times 3}$ contains three weights $m_{sub}$, $m_{com}$ and $m_{reg}$ , denoting the user's intent toward substitutes, complements, and other regular items, respectively. In \modelname, RecTool memory is optimized by tool comparison and ranking comparison modules, detailed in Sec.~\ref{sec:tool_compare} and~\ref{sec:rank_compare}. Intent memory is optimized by a intent discrimination module in Sec.~\ref{sec:intent_compare}.

\subsection{Substitute and Complement Generation}~\label{sec:scgen}
A key challenge in enhancing personalization through substitution and complementarity is that such relationships are usually implicit in most datasets. Annotating these pairwise relationships with LLMs across a large number of items demands substantial computational resource, as it involves $\mathcal{O}(|\mathcal{I}|^2)$ LLM inference calls. To mitigate computational overhead, \modelname is designed to refine the top-ranked items from recommendation tools based on the substitutes and complements of the user’s latest preferences. The first step in this process involves generating potential substitutes and complements based on the user’s historical behaviors. Specifically, we prompt the LLM to generate a list of substitutes and a list of complements corresponding to the user’s recent interests:
\begin{align}
    y_{\text{gen}}^{sub} &= \text{LLM}_\text{gen}^{sub}(\text{desc}(\mathbf{s}_{[-(k+c):-k]})) \label{eq:gen_sub}\\
    y_{\text{gen}}^{com} &= \text{LLM}_\text{gen}^{com}(\text{desc}(\mathbf{s}_{[-(k+c):-k]})), \label{eq:gen_com}
\end{align}
where $\mathbf{s}_{[-(k+c):-k]}$ denotes the most recent $c$ items visible to the recommendation tools in the sequence. The textual outputs $y_{\text{gen}}^{sub}$ and $y_{\text{gen}}^{com}$ represent the generated lists of substitutes and complements, each of length $c$. The functions $\text{LLM}_\text{gen}^{sub}(\cdot)$ and $\text{LLM}_\text{gen}^{com}(\cdot)$ denote the prompt functions for generating substitutes and complements, respectively.
From a complexity perspective, we reduce the LLM inference calls from $\mathcal{O}(|\mathcal{I}|^2)$ for pairwise annotation to $\mathcal{O}(|\mathcal{U}|)$ for personalized intent generation. This reduction substantially enhances scalability in real-world scenarios, such as online grocery involving millions of items.

\subsection{Personalized Tool Selection}~\label{sec_tool_select}
\subsubsection{LLM-based Tool Comparison}~\label{sec:tool_compare}
Different users exhibit distinct behavior patterns and therefore benefit from different recommendation strategies. For instance, users who read books like \textit{Harry Potter} often proceed with its sequels, making sequential recommendation models particularly effective. After finishing all the sequels, users may be interested in other fictions read by similar users, where collaborative filtering proves more suitable. To enable personalized tool selection for each user, \modelname incorporates a tool comparison module that performs a holistic evaluation across multiple recommendation tools. In this module, the user agent is tasked with identifying the best tool based on recommended results of tools and ground-truth user preference:
\begin{equation}~\label{eq:tool_cmpr}
    \mathbf{z}_{\text{cpr}} = \text{LLM}_{\text{cpr}}(\{\text{desc}(f_\text{top}(\hat{\mathbf{r}}_{t}, k_{\text{cpr}}))|t\in \mathcal{T}\}, \text{desc}(\mathbf{s}_{[-k:]})),
\end{equation}
where the output $\mathbf{z}_{\text{cpr}}$ consists of $|\mathcal{T}|$ binary indicators, each indicating whether tool $t$ provides the most relevant recommendation to the user's preference among all tools in $\mathcal{T}$. For instance, $z_{\text{cpr}}^{G} = 1$ indicates that the LLM has selected the graph-based tool as the best match to the user’s preferences.
We use $f_\text{top}(\hat{\mathbf{r}}_{t}, k_{\text{cpr}})$ to represent the sequence of top-$k_{\text{cpr}}$ items ranked by recommendation tool $t$. Subsequence $\mathbf{s}_{[-k:]}$ refers to the most recent $k$ user-interacted items, which are withheld from training of recommendation tools and used as ground-truth signals to optimize the agent's decision-making. The function $\text{LLM}_\text{cpr}(\cdot)$ is used to prompt the LLM to compare the input sequences and select the tool most aligned with the user's preferences.
After the LLM identifies the tool that best aligns with the user’s recent preferences, the tool weight $m_{t}\in \{m_G, m_S, m_M\}$ in RecTool memory is updated accordingly:
\begin{equation}\label{eq:toolcpr}
    m_{t} \leftarrow  m_{t} + \alpha\cdot z_{\text{cpr}}^{t}, 
\end{equation}
where $\alpha$ is the learning rate to update RecTool memory by this module. 
This tool comparison remains robust even when multiple tools appear effective. It addresses two key challenges: LLMs tend to agree with any reasonable recommendation, and single-domain datasets often make all outputs appear relevant.

\subsubsection{Ranking Comparison and Aggregation}~\label{sec:rank_compare}
Besides the LLM-based tool comparison, we also compare the ranking performance based on the ranks of the most recent $k$ items in each tool:
\begin{equation}~\label{eq:rankcmpr}
    m_{t} \leftarrow  m_{t} + \beta\sum_{\{r^j_{t}|i_j\in\mathbf{s}_{[-k:]}\}}\frac{(r^j_{t})^{-1}}{\sum_{t'\in\mathcal{T}}(r^j_{t'})^{-1}},
\end{equation}
where $r^j_{t}$ is the rank of the $j$-th item from tool $t$.
This quantitative comparison yields a behavior-grounded signal reflecting how well each tool generalizes to unseen user preferences in $\mathbf{s}_{[-k:]}$. 
To obtain a comprehensive ranked list from all tools, a weighted sum is computed based on the updated RecTool memory:
\begin{equation}~\label{eq:aggrank}
    \hat{\mathbf{r}} = \sum_{t\in\mathcal{T}}m_t\hat{\mathbf{r}}_t = m_G\hat{\mathbf{r}}_{G} +m_S\hat{\mathbf{r}}_{S} + m_M\hat{\mathbf{r}}_{M},
\end{equation}
where $\hat{\mathbf{r}}\in\mathbb{R}^{1\times|\mathcal{I}|}$ is the aggregated ranked item list. This simple yet effective rank-level ensemble strategy allows the agent to adaptively combine tools based on user-specific preferences. It also remains flexible and can be replaced by more complex list-wise combination mechanisms based on backpropagation. We implement and compare different ensemble strategies for ranking comparison and aggregation in Sec.~\ref{sec:ensemble}.

\subsection{Intent Discrimination}~\label{sec:intent_compare}
Understanding whether a user’s next interaction is driven by substitutes, complements, or general interest is crucial for tailoring recommendations that match their intent. Traditional ID-based recommenders struggle to distinguish these patterns, as they require nuanced semantic and functional reasoning that goes beyond co-occurrence statistics. With reasoning ability of LLMs, user agents can allocate attention to the most relevant pathway while ensuring coverage of diverse user behaviors, ultimately improving ranking robustness and relevance.
\subsubsection{Substitution and Complementarity}
Based on implicit item-item relationships, we first distinguish the user's potential intent between substitutes and complements. Given the substitute list $y_{\text{gen}}^{sub}$ and the complement list $y_{\text{gen}}^{com}$ generated in Sec.~\ref{sec:scgen}, we prompt the LLM to access the user's interest in substitute or complements by comparing the generated lists with the ground-truth preferences:
\begin{equation}~\label{eq:scdcm}
     \mathbf{z}_{\text{dcm}} = \text{LLM}_\text{dcm}(y_{\text{gen}}^{sub}, y_{\text{gen}}^{com}, \text{desc}(\mathbf{s}_{[-k:]})),
\end{equation}
where the output $\mathbf{z}_{\text{dcm}}$ consists of two binary indicators $z_{\text{dcm}}^{sub}$ and $z_{\text{dcm}}^{sub}$, which sum to 1. Each indicator reflects whether the corresponding list exhibits greater semantic and functional alignment with the user's recent interactions. The function $\text{LLM}_\text{dcm}(\cdot)$ prompts the LLM to identify whether the substitute or complement list better matches the user’s recent preferences.

\subsubsection{General Interest}
In addition, a separate module is employed to assess the user's general interest in items other than substitutes or complements, based on the user’s historical behavior:
\begin{equation}\label{eq:reglearn}
     z_{\text{dcm}}^{reg} = \text{LLM}_\text{dcm}^{reg}(\text{desc}(\mathbf{s})),
\end{equation}
where the output $z_{\text{dcm}}^{reg}$ of prompt function $\text{LLM}_\text{dcm}^{reg}(\cdot)$ is a binary indicator that equals 1 if the ground-truth user preferences do not exhibit clear substitute or complement patterns otherwise 0. This additional assessment ensures coverage of a broader spectrum of user behaviors, particularly in scenarios where the user’s preferences are not driven by substitution or complementarity. Identifying such general interest prevents the agent from overfitting to relational reasoning. This mitigates the risk of forcing item-item relational interpretations where they do not apply.

\subsubsection{General Reranking}\label{sec:genrerank}
To account for general interests of users, a reranking module is introduced to refine the recommendations based on the semantic similarity between the candidate items and the users' general preferences. For each user, the general preferences have been summarized from historical items as user profile in Sec.~\ref{sec:userprof}. Then the LLM is prompted to rerank the top items based on the user profile $y_{\text{prof}}$, denoted as function $\text{LLM}^{reg}_{\text{rank}}(\cdot)$:
\begin{equation}~\label{eq:gen_rerank}
    y_{\text{rank}}^{reg} = \text{LLM}^{reg}_{\text{rank}}(f_\text{top}(\hat{\mathbf{r}},k'), \text{desc}(f_\text{top}(\hat{\mathbf{r}},k')),y_{\text{prof}}),
\end{equation}
where $f_\text{top}(\hat{\mathbf{r}},k')$ is the top $k'$ items in the aggregated ranked list $\hat{\mathbf{r}}$, and $y_{\text{rank}}^{reg}$ is the reranked list containing $k'$ item IDs.

\subsubsection{Intent Memory Update}
Similar to RecTool memory update in Eq.~\ref{eq:toolcpr}, the corresponding intent $m_{int}\in \{m_{sub}, m_{com}, m_{reg}\}$ is updated based on $z_{\text{dcm}}^{int}\in \{z_{\text{dcm}}^{sub}, z_{\text{dcm}}^{com}, z_{\text{dcm}}^{reg}\}$ by a learning rate $\gamma$:
\begin{equation}\label{eq:intentdcm}
    m_{int} \leftarrow  m_{int} + \gamma\cdot z_{\text{dcm}}^{int}.
\end{equation}
This intent discrimination module equips the agent with a nuanced understanding of user preferences, enabling it to adaptively allocate attention to different reasoning modes and thereby guide downstream reranking in a more personalized manner.

\subsection{Dual S\&C Reranking}\label{sec:dualrerank}
A dual substitute and complement (S\&C) reranking module is proposed to comprehensively refine the recommendation results from tools in \modelname. First, to leverage substitute and complement relation reasoning to enhance recommendation via world knowledge of the LLM, we design two independent reranking modules to refine the top-ranked items based on their semantic similarity with the candidate substitutes and complements for a user:
\begin{align}
    y_{\text{rank}}^{sub} &= \text{LLM}_{\text{rank}}(f_\text{top}(\hat{\mathbf{r}},k'), \text{desc}(f_\text{top}(\hat{\mathbf{r}},k')),y_{\text{gen}}^{sub}) \label{eq:rerank_sub}\\
    y_{\text{rank}}^{com} &= \text{LLM}_{\text{rank}}(f_\text{top}(\hat{\mathbf{r}},k'), \text{desc}(f_\text{top}(\hat{\mathbf{r}},k')),y_{\text{gen}}^{com}), \label{eq:rerank_com}
\end{align}
where output $y_{\text{rank}}^{sub}$ and $y_{\text{rank}}^{com}$ are the reranked lists containing the same $k'$ items from $f_\text{top}(\hat{\mathbf{r}},k')$. The LLM is prompted to rerank the top items from the aggregated ranked list based on their similarity to candidate substitutes and complements, denoted as $\text{LLM}_{\text{rank}}(\cdot)$. 

Although the expected output of the reranked list is a sequence of exactly $k'$ item IDs selected from the ID-only list $f_\text{top}(\hat{\mathbf{r}},k')$, the LLM may hallucinate by generating invalid outputs such as non-existent item IDs or irrelevant text. To address this issue, we apply a rule-based hallucination filtering mechanism to ensure output validity. Specifically, only valid item IDs in the item corpus are retained in the reranked list. If the resulting list contains fewer than $k'$ items, the remaining valid IDs from $f_\text{top}(\hat{\mathbf{r}},k')$ are appended to the end of the valid list in original order. This filtering mechanism is also applied in the aforementioned general reranking.
% Compared to prior approaches that rely on carefully crafted query prompts or self-reflection mechanisms to mitigate hallucination~\cite{agentcf, agent4rec, iagent}, 
This strategy completely eliminates the impact of hallucination on the ranking task, while incurring minimal computational overhead and requiring minimal reliance on the LLM’s instruction-following capability.

Subsequently, to account for individual variation in user intent, we personalize the combination of the two reranked ID lists using a ranking fusion function $f_{\text{fuse}}(\cdot)$. This function takes the two ID lists and the corresponding intent weights as input:
\begin{equation}\label{eq:fuse}
    \mathbf{\hat{r}}^{dual} = f_{\text{fuse}}((m_{sub}, y_{\text{rank}}^{sub}), ( m_{com}, y_{\text{rank}}^{com})),
\end{equation}
where the ranking score $\hat{r}_{i}^{dual}$ of each item $i$ in this output list $\mathbf{\hat{r}}^{dual}$ is computed as a weighted sum of the item's positions in the two input lists:
\begin{equation}
    \hat{r}_{i}^{dual} = m_{sub}\cdot\text{index}(i, y_{\text{rank}}^{sub}) + m_{com}\cdot\text{index}(i, y_{\text{rank}}^{com}), 
\end{equation}
where $\text{index}(i,\cdot)$ returns the rank position of item $i$ within a given list. This personalized fusion strategy allows the agent to balance substitute-based and complement-based reasoning in accordance with the inferred user intent, without incurring additional LLM inference overhead.
The output list $\mathbf{\hat{r}}^{dual}$ is further fused with the item list from general reranking:
\begin{equation}\label{eq:regfuse}
    \mathbf{\hat{r}}^{final} = f_{\text{fuse}}((1, y_{\text{rank}}^{dual}), (m_{reg}, y_{\text{rank}}^{reg})),
\end{equation}
where $y_{\text{rank}}^{dual}$ denotes the item list sorted by ranking scores in $\mathbf{\hat{r}}^{dual}$. The final recommendation output consists of the top-$k'$ items ranked by their scores in $\mathbf{\hat{r}}^{final}$. By incorporating the broader user preference profile as a form of regularization in the reranking process, the agent achieves a more robust recommendation strategy, particularly when relational cues among items are weak or absent.

\section{Experiment}
Through our extensive empirical analysis, we aim to address the following research questions. 
\textbf{(RQ1)}~How does \modelname perform in full-ranking recommendation compared to other baselines?\linebreak[4]
\noindent \textbf{(RQ2)}~Does incorporating LLM-based reasoning modules improve the semantic alignment of recommended items with user intent?
\textbf{(RQ3)}~How do different ranking aggregation strategies affect the performance of \modelname?
\textbf{(RQ4)}~What is the individual contribution of each LLM-based module in \modelname?
% \begin{itemize}[left=0pt]
%     \item \textbf{RQ1:} How does \modelname perform in full-ranking recommendation compared to other baselines?
%     \item \textbf{RQ2:} Does incorporating LLM-based reasoning modules improve the semantic alignment of recommended items with user intent?
%     \item \textbf{RQ3:} How do different ranking aggregation strategies affect the performance of \modelname?
%     \item \textbf{RQ4:} What is the individual contribution of each LLM-based module in \modelname?
% \end{itemize}
\subsection{Experiment Settings}
\subsubsection{Datasets}
We conduct experiments on three datasets: Instacart~\cite{instacart}, %\footnote{https://www.kaggle.com/yasserh/datasets}, 
Electronics and Sports~\cite{bonab2021crossmarket}. %\footnote{https://xmrec.github.io/data/us/}. 
% Instacart is a public dataset containing transactional data from the Instacart app, primarily focused on grocery items for basket recommendation. 
Electronics and Sports are two sub-datasets from public XMarket dataset~\cite{xmarket}.
% Electronics contains user history of purchasing electronics on Amazon. 
% Sports contains user interactions with items under category of sports and outdoors on Amazon. 
The statistics are summarized in Table~\ref{tab:datasets}. 
\begin{table}
  \caption{Dataset statistics}
 \vspace{-0.35cm}
  \centering
  \label{tab:datasets}\resizebox{0.35\textwidth}{!}{
  \begin{tabular}{l c c c}
        \toprule
        % \hline
        % \hline
        \textbf{Dataset} & Instacart & Electronics & Sports \\
        \midrule
        \textbf{User \#} & 6,987 & 4,966 & 10,873\\

        \textbf{Item \#} & 2,108 & 14,635 & 4,306\\

        \textbf{Interaction \#} & 5,105,339 & 91,819 & 55,990\\
        \textbf{Density} & 34.663\% & 0.126\% & 0.120\% \\
        \bottomrule
        % \hline
  \end{tabular}}
 \vspace{-0.4cm}
\end{table}
\subsubsection{Tool selection}\label{sec:toolselect}
In this work, three pretrained recommendation models are used as tools within \modelname: 
\textbf{(1) LightGCN}~\cite{LightGCN}, \textbf{(2) SASRec}~\cite{sasrec}, and \textbf{(3) SimpleX}~\cite{simplex}.
% \begin{enumerate}[left=0pt, label=\textbf{(\arabic*)}]
%     \item \textbf{LightGCN}~\cite{LightGCN}. A graph-based collaborative filtering model that simplifies traditional graph convolution network by removing feature transformations and nonlinearities.
%     % Same as matrix factorization (MF), its space complexity is $O((|\mathcal{I}|+|\mathcal{U}|)\cdot d)$ and time complexity is $O(L\cdot|E|\cdot d)$, where $L$ is the number of graph convolution layers, $|E|$ is the number of interactions and $d$ is the embedding size.
%     \item \textbf{SASRec}~\cite{sasrec}.  A sequential recommendation model based on Transformer, which captures the user's dynamic preferences by modeling the dependencies between historical items.
%     % Its space complexity is $O(|I|\cdot d + n\cdot d + d)$ and time complexity is $O(|\mathcal{U}|\cdot (n^2\cdot d+n\cdot d^2))$ per taining epoch, where $n$ is the length of behavior sequences.
%     \item \textbf{SimpleX}~\cite{simplex}. A MF-based model where a margin is applied to filter uninformative negative samples in the loss.
%     % It has the same space complexity with MF. The time complexity is $O(|E|\cdot (n+\overline{n})\cdot d)$ where $\overline{n}$ is the number of negative samples per interaction. 
% \end{enumerate}
The simplicity of these tools allows us to validate the effectiveness of our framework without relying on complex architectures, aligning with our goal of maintaining a model-agnostic design.
\subsubsection{Baselines}
Besides the individual recommendation tools, we compare \modelname with eight other baselines.
First, other representative MF-based, diffusion-based, and sequential models are used as tradition recommendation baselines:
\begin{enumerate}[left=0pt, label=\textbf{(\arabic*)}, start=4]
    \item \textbf{ENMF}~\cite{enmf} introduces mathematical optimization that enable efficient full-data MF without negative sampling.
    \item \textbf{DiffRec}~\cite{diffrec} is a diffusion-based recommendation model that generates user interactions in a denoising manner.
    \item \textbf{FEARec}~\cite{fearec} enhances self-attention with frequency-aware mechanisms to capture high-frequency signals and periodic patterns for sequential recommendation.
\end{enumerate}
Both tools and recommendation baselines are pretrained on the first $(n-k)$ items of each user’s interaction sequence $\mathbf{s}_{[:-k]}$, following the standard training protocol of sequential recommendation for fair performance comparison.
% Since \modelname is complementary to the existing full-ranking models and aims to enhance their recommendations via the reasoning capabilities and world knowledge of LLMs, further comparison with other individual recommendation models is left for future work. 
Following the previous works~\cite{agentcf, iagent}, we also compare \modelname with language-based approaches:
\begin{enumerate}[left=0pt, label=\textbf{(\arabic*)}, start=7]
    \item \textbf{BM25}~\cite{bm25} ranks candidates according to their textual similarity with user historical interactions.
    \item \textbf{LLMRanker}~\cite{llmrank} uses the LLM as a zero-shot ranker, considering user sequential interaction histories as conditions.
\end{enumerate}
In addition to these language-based baselines, 
we compare \modelname against recommendation tools integrated with Retrieval-Augmented Generation (RAG). 
Specifically, the LLM is prompted to generate the top 20 items from the 50 candidates retrieved by each tool, 
conditioned on the user’s behavior sequence. 
The same hallucination filtering mechanism from Sec.~\ref{sec:dualrerank} is applied for fair comparison. 
These baselines are represented as \textbf{(9)-(11) \boldmath RAG$_{\mathit{G/S/M}}$} for LightGCN, SASRec and SimpleX, respectively. 
Since most prior LLM-agent works~\cite{agentcf,recmind,interecagent,agent4rec,recagent,iagent} are unable to rank all items in datasets, we exclude them from the comparison table instead of reporting trivially low full-ranking performance.
\subsubsection{Evaluation Metrics}
Recall@\{10,20\} and NDCG@\{10,20\} are adopted for ranking performance evaluation. Additionally, we propose a novel metric that measures the semantic vicinity of the predicted item list to the ground-truth item in the following Sec.~\ref{sec:vdcg}.

\subsection{Semantic Vicinity Evaluation: VDCG}~\label{sec:vdcg}
Unlike ID-based recommenders, language-based methods prioritize semantic similarity over behavioral patterns. As a result, the item lists they generate may better reflect a user’s interests, even if the exact target items are absent. For example, for a user interested in a black baseball helmet, a list of baseball helmets in other colors is clearly more relevant than a list of bicycle helmets, despite both yielding zero recall and NDCG. To better capture the semantic vicinity to the user intent, we propose a new LLM-based evaluation metric, Vicinity-DCG (VDCG), which jointly assesses the semantic alignment and ranking correctness of the predicted item list.

Specifically, we prompt the LLM to rate how well each item in the recommended list aligns with the ground-truth item in terms of relevance, usefulness, and user interest. The rating scale ranges from 0 to 9, where 0 indicates an entirely unrelated item, and 9 represents a perfect semantic match between the item descriptions.
These ratings are treated as relevance scores to compute the DCG~\cite{dcg} for the list. For normalization, we divide by a customized ideal DCG, calculated from a synthetic ideal list of scores: $[p, p/2, ..., p/2^l]$, where $p$ is the maximum relevance score minus 1, and 
$l$ is the length of the list. This ideal score distribution reflects the empirical tendency of the LLM to assign higher scores to only a few semantic-related items, following an approximately exponential decay. VDCG thus captures both the semantic vicinity and ranking quality of the predicted list with respect to the user's true intent.

\subsection{RQ1: Ranking Performance Comparison}

\begin{table*}[]\caption{Overall performance comparison. The best and second-best performances are in boldface and underlined.}\label{tab:overall}
\vspace{-3mm}
\resizebox{0.9\textwidth}{!}{%
\begin{tabular}{l rrrr @{\hspace{1.5em}} rrrr @{\hspace{1.5em}} rrrr}
\toprule
%Dataset
& \multicolumn{4}{c}{Instacart}
& \multicolumn{4}{c}{Electronics}
& \multicolumn{4}{c}{Sports} \\
\cmidrule(lr){2-5}\cmidrule(lr){6-9}\cmidrule(lr){10-13}
%Metric
& R@10 & R@20 & N@10 & N@20
& R@10 & R@20 & N@10 & N@20
& R@10 & R@20 & N@10 & N@20 \\
\midrule
LightGCN  & 0.0500 & 0.1188 & 0.0221 & 0.0398 & 0.1188 & 0.1313 & 0.0734 & 0.0765 & \uline{0.1438} & 0.1625 & 0.0997 & 0.1040 \\
SASRec    & 0.0875 & 0.1500 & 0.0395 & 0.0552 & 0.1188 & 0.1313 & 0.0635 & 0.0667 & 0.1000         & 0.1438 & 0.0554 & 0.0667 \\
SimpleX   & 0.0313 & 0.0500 & 0.0180 & 0.0226 & 0.1000 & 0.1063 & 0.0761 & 0.0776 & 0.1375         & 0.1563 & 0.0951 & 0.0999 \\
\midrule
ENMF      & 0.1125 & 0.1750 & 0.0711 & 0.0797 & 0.1250 & 0.1312 & 0.0881 & 0.0897 & 0.1375         & 0.1438 & \uline{0.1172} & \uline{0.1187} \\
DiffRec   & \uline{0.1188} & 0.1563 & \uline{0.0773} & 0.0806 & \uline{0.1562} & \uline{0.1688} & \uline{0.0971} & \uline{0.0981} & 0.1125 & 0.1188 & 0.0875 & 0.0890 \\
FEARec    & 0.1063 & \uline{0.1875} & 0.0639 & \uline{0.0845} & 0.1250 & 0.1625 & 0.0627 & 0.0726 & 0.1313 & 0.1563 & 0.0732 & 0.0797 \\
\midrule
BM25      & 0.0625 & 0.0875 & 0.0310 & 0.0370 & 0.1000 & 0.1313 & 0.0432 & 0.0514 & 0.1188 & 0.1250 & 0.0536 & 0.0552 \\
LLMRank   & 0.0750 & 0.1000 & 0.0373 & 0.0436 & 0.1000 & 0.1250 & 0.0503 & 0.0565 & 0.0938 & 0.1063 & 0.0458 & 0.0488 \\
\midrule
RAG$_G$   & 0.1125 & 0.1563 & 0.0480 & 0.0594 & 0.1125 & 0.1438 & 0.0715 & 0.0748 & 0.1125 & 0.1188 & 0.0750 & 0.0768 \\
RAG$_S$   & 0.0688 & 0.1063 & 0.0312 & 0.0405 & 0.1063 & 0.1188 & 0.0575 & 0.0605 & 0.1375 & \uline{0.1625} & 0.0852 & 0.0913 \\
RAG$_M$   & 0.0750 & 0.1125 & 0.0335 & 0.0427 & 0.0875 & 0.1250 & 0.0517 & 0.0613 & 0.1125 & 0.1375 & 0.0823 & 0.0871 \\
\midrule
\modelname & \textbf{0.1563} & \textbf{0.2000} & \textbf{0.0796} & \textbf{0.0907}
          & \textbf{0.1688} & \textbf{0.2000} & \textbf{0.1180} & \textbf{0.1260}
          & \textbf{0.1750} & \textbf{0.2063} & \textbf{0.1247} & \textbf{0.1326} \\
Improv.   & 33.50\% & 6.67\% & 2.98\% & 7.34\%
          & 8.07\%  & 18.48\% & 21.52\% & 28.44\%
          & 21.67\% & 26.95\% & 6.40\% & 11.71\% \\
\bottomrule
\end{tabular}%
}
\vspace{-3mm}
\end{table*}

Performance comparison between \modelname and other baselines are shown in Table~\ref{tab:overall}. We have the following observations:
\begin{itemize}[left=0pt]
    \item \modelname exhibits superior performance across all datasets. Even compared with more advanced and complex recommendation methods such as DiffRec, it has up to $33.5\%$ improvement in recall and up to $28.4\%$ improvement in NDCG. When compared to its own base recommendation tools such as SASRec, \modelname consistently achieves at least $33.3\%$ improvement across all datasets. Since \modelname's performance is inherently dependent on the underlying tools, we expect greater performance gains when integrated with more advanced recommenders. We leave the exploration of stronger underlying tools in future works, as any resulting performance gain would not stem from our agent design and thus falls outside the scope of this study.
    \item Language-only methods, such as BM25 and LLMRank, underperform compared to all recommendation baselines in most scenarios. This performance gap arises from their inability to effectively capture collaborative filtering signals or sequential behavioral patterns, both of which are essential for modeling personalized user preferences in large-scale item spaces. Lacking the ability to leverage historical interaction structures, these methods rely solely on surface-level text matching or general world knowledge, which limits their performance in full-ranking recommendation.
    \item Compared to \modelname, RAG-based methods such as RAG$_S$, fail to consistently improve recommendation performance on all the datasets through the world knowledge of LLMs. This is due to two main reasons. First, relying on a single recommender as the retrieval model leads the LLM to generate top items from a potentially biased candidate set, even if the set is larger than that used in \modelname. Second, the behavior sequences fed to the LLM consist of item descriptions. These descriptions usually contain noises, such as abbreviations or symbols, that carry little to no preference-relevant information. In contrast, \modelname reranks a smaller but higher-confidence candidate set aggregated from multiple recommendation tools and reasons over a summarized user profile with inferred intent, enabling more comprehensive and nuanced recommendations.
\end{itemize}

\subsection{RQ2: VDCG Comparison}
The VDCG comparison is presented in Table~\ref{tab:vdcg}. Compared to the ranking-based evaluation results in Table~\ref{tab:overall}, RAG-based methods demonstrate a much more pronounced advantage in VDCG over the base recommenders. Given their overall lower recall, this advantage primarily stems from the stronger semantic vicinity of their recommended item lists to user interests. This demonstrates that the world knowledge embedded in LLMs can be effectively harnessed to generate semantically meaningful recommendations aligned with user intent, which is especially valuable when the target item is difficult to predict from historical behavior.
Similarly, language-only methods such as BM25 and LLMRank achieve competitive VDCG scores despite poor recall and NDCG performance, especially in the Electronics and Sports where item descriptions tend to be longer and more semantically distinguishable. In contrast, the relatively low VDCG scores of interaction-based recommenders in these datasets reveal a limitation not captured by standard recall or NDCG metrics. In real-world e-commercial services, a lack of semantic relevance in recommended lists may undermine user trust more severely than occasional misses. Addressing this gap, \modelname actively reasons over implicit item-item relations, resulting in the highest overall semantic and ranking alignment, as reflected by its VDCG.

\begin{figure}[]
    \begin{subfigure}{0.163\textwidth}
    \includegraphics[width=\textwidth]{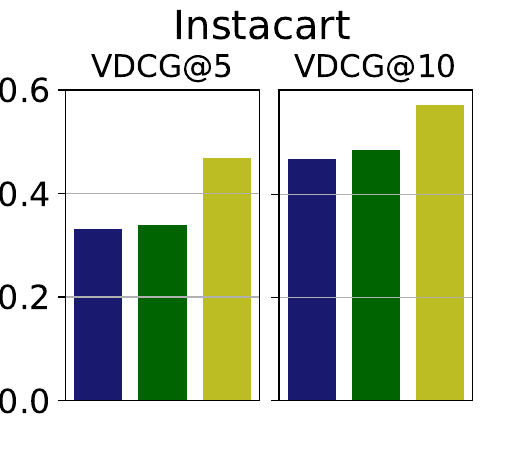}
    \end{subfigure}
    \hspace{-3mm}
    % \hfill
    \begin{subfigure}{0.163\textwidth}
    \includegraphics[width=\textwidth]{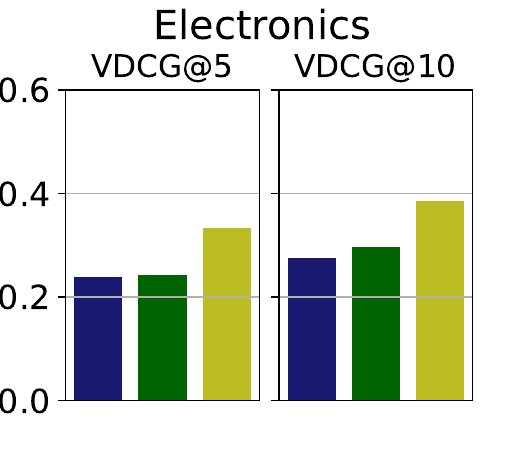}
    \end{subfigure}
    \hspace{-3mm}
    % \hfill
    \begin{subfigure}{0.163\textwidth}
    \includegraphics[width=\textwidth]{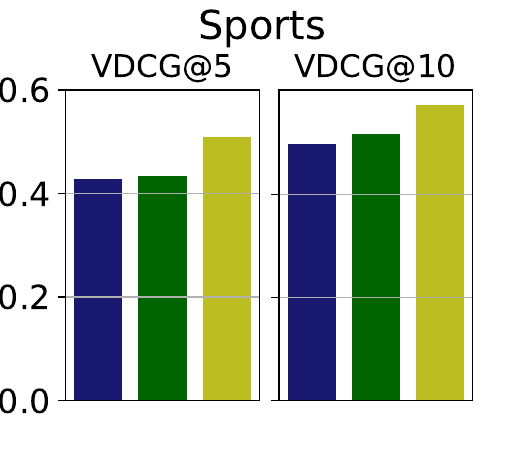}
    \end{subfigure}
    \begin{subfigure}{0.45\textwidth}
    \centering
   \vspace{-2mm}
    \includegraphics[width=0.9\textwidth]{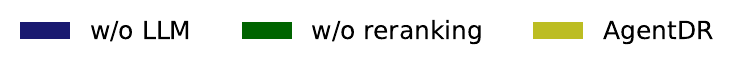}
    \end{subfigure}
    \vspace{-4mm}
        \caption{Performance of \modelname on VDCG without LLM-based modules. The bars of w/o LLM refer to \modelname without both reranking and tool comparison on three datasets.}
  \label{fig:vdcg}
\end{figure}

In Fig.~\ref{fig:vdcg}, we further examine the contribution of the LLM-based reasoning modules to VDCG by progressively removing them from the ranking refinement and agent optimization stages. The VDCG gap between \modelname with and without the reranking modules highlights their importance in enhancing the semantic alignment between recommended item lists and user preferences. Notably, tool comparison consistently improves VDCG, as it increases the weight of the recommendation tool that produces the most semantically relevant item list during agent optimization.

\begin{table}
\caption{Performance comparison on VDCG, which jointly evaluates semantic and ordering correctness.}
\label{tab:vdcg}
\vspace{-3mm}
\resizebox{0.475\textwidth}{!}{%
\begin{tabular}{l cc @{\hspace{1.2em}} cc @{\hspace{1.2em}} cc}
\toprule
%Dataset
& \multicolumn{2}{c}{Instacart}
& \multicolumn{2}{c}{Electronics}
& \multicolumn{2}{c}{Sports} \\
\cmidrule(lr){2-3}\cmidrule(lr){4-5}\cmidrule(lr){6-7}
%Metric
& V@5 & V@10
& V@5 & V@10
& V@5 & V@10 \\
\midrule

BM25      & 0.2409 & 0.3220 & 0.2547 & 0.3188 & 0.5030 & \uline{0.6075} \\
LLMRank   & 0.2529 & 0.3224 & 0.2180 & 0.2932 & 0.4351 & 0.5341 \\
\midrule

LightGCN  & 0.3144 & 0.4039 & 0.1736 & 0.2344 & 0.4541 & 0.5702 \\
SASRec    & 0.3143 & 0.4037 & 0.1748 & 0.2377 & 0.3898 & 0.4847 \\
SimpleX   & 0.3091 & 0.3984 & 0.1735 & 0.2341 & 0.4551 & 0.5884 \\
\midrule

RAG$_G$   & \uline{0.3363} & \uline{0.4537} & \uline{0.2656} & \uline{0.3338} & 0.4119 & 0.5220 \\
RAG$_S$   & 0.3300 & 0.4496 & 0.2569 & 0.3257 & \textbf{0.5095} & \textbf{0.6094} \\
RAG$_M$   & 0.3306 & 0.4460 & 0.2100 & 0.2874 & 0.4670 & 0.5768 \\
\midrule

\modelname & \textbf{0.4683} & \textbf{0.5743} & \textbf{0.3334} & \textbf{0.3846} & \uline{0.5085} & 0.5743 \\
\midrule

Improv.   & 39.25\% & 26.53\% & 25.90\% & 15.21\% & -0.20\% & -5.76\% \\
\bottomrule
\end{tabular}%
}
\vspace{-3mm}
\end{table}

% previous version
% \begin{table}[]\caption{Performance comparison on the proposed VDCG which jointly evaluate the
% semantic and ordering correctness.}\label{tab:vdcg}
% \vspace{-3mm}
% \resizebox{0.475\textwidth}{!}{
% \begin{tabular}{l|cc|cc|cc}
% \hline
% \multicolumn{1}{c|}{Dataset} & \multicolumn{2}{c|}{Instacart}    & \multicolumn{2}{c|}{Electronics}       & \multicolumn{2}{c}{Sports}     \\ \hline
% Metric         & V@5 & V@10             & V@5             & V@10             & V@5             & V@10             \\ \hline
% BM25 & 0.2409 & 0.3220 & 0.2547 & 0.3188 & 0.5030	& \uline{0.6075} \\
% LLMRank & 0.2529 & 0.3224 & 0.2180 & 0.2932 & 0.4351 & 0.5341 \\ \hline
% LightGCN & 0.3144 & 0.4039 & 0.1736 & 0.2344 & 0.4541 &0.5702 \\
% SASRec  & 0.3143 & 0.4037 & 0.1748 & 0.2377 & 0.3898 & 0.4847 \\
% SimpleX & 0.3091 & 0.3984 & 0.1735 & 0.2341 & 0.4551 & 0.5884 \\ \hline
% RAG$_{G}$  & \uline{0.3363} & \uline{0.4537} & \uline{0.2656} & \uline{0.3338} & 0.4119 & 0.5220 \\
% RAG$_{S}$ & 0.3300 & 0.4496 & 0.2569 & 0.3257 & \textbf{0.5095} & \textbf{0.6094} \\
% RAG$_{M}$ & 0.3306 & 0.4460 & 0.2100 & 0.2874 & 0.4670 & 0.5768 \\ \hline
% \modelname  & \textbf{0.4683} & \textbf{0.5743} & \textbf{0.3334} & \textbf{0.3846} & \uline{0.5085} & 0.5743\\ \hline
% Improv. & \multicolumn{1}{l}{39.25\%} & \multicolumn{1}{l|}{26.53\%} & \multicolumn{1}{l}{25.90\%} & \multicolumn{1}{l|}{15.21\%} & \multicolumn{1}{l}{-0.20\%} & \multicolumn{1}{l}{-5.76\%} 
% \\ \hline
% \end{tabular}
% }
% \vspace{-3mm}
% \end{table}

\subsection{RQ3: Aggregation Strategy Analysis}
~\label{sec:ensemble}

\vspace{-0.3cm}
\subsubsection{Setup} The ranking comparison in Eq.~\ref{eq:rankcmpr} is a simple quantitative reflection mechanism to adjust RecTool weights for personalized tool suitability.  This component is flexible and can be replaced with alternative learning-based ensemble methods.  For instance, we can train a learnable vector $\mathbf{v}\in \mathbb{R}^{1 \times |\mathcal{T}|}$ as adaptive tool weights for all users. This learnable vector is then list-wise multiplied by the ranking score lists $\hat{\mathbf{r}}_\mathcal{T}\in \mathbb{R}^{|\mathcal{T}| \times |\mathcal{I}|}$ of each user to obtain the aggregated ranking score list $\hat{\mathbf{r}}'\in\mathbb{R}^{1\times|\mathcal{I}|}$. Supervised training is performed using ground-truth target items in training set, with a hinge loss as the objective. To integrate this with the original framework, we normalize this $\hat{\mathbf{r}}'$ and $\hat{\mathbf{r}}$ from Eq.~\ref{eq:aggrank} and add them as an updated $\hat{\mathbf{r}}$ for the following reranking.

\subsubsection{Results} Table~\ref{tab:ensemble} reports the performance of various ensemble strategies, both with and without all LLM-based reasoning modules from \modelname. The original ranking comparison mechanism and the alternative linear model described above are denoted as \textit{RC} and \textit{LR}, respectively. Additionally, we explore MLP with and without attention mechanisms to capture inter-tool dependencies, as alternatives to the original ranking comparison. These variants are denoted as \textit{Att} and \textit{MLP} in the table. 
The results demonstrate that the proposed reasoning modules consistently provide significant benefits across different ensemble strategies. Although \modelname equipped with the simplest ranking comparison mechanism already surpasses most baselines in Table~\ref{tab:overall}, substituting it with more expressive ensemble methods yields up to a 13.56$\%$ improvement in NDCG. This performance gain comes at the cost of increased computational complexity during backpropagation. Among the evaluated methods, the MLP with an attention mechanism achieves the best overall performance when integrated into \modelname, which has the the highest number of learnable parameters. We leave the exploration of potential improvement from integrating more advanced ensemble methods in the future work.

\begin{table}
\caption{Performance of different ensemble methods with and without all LLM reasoning modules from \modelname.}
\label{tab:ensemble}
\vspace{-3mm}
\resizebox{0.46\textwidth}{!}{%
\begin{tabular}{l cc @{\hspace{1.2em}} cc @{\hspace{1.2em}} cc}
\toprule
%Dataset
& \multicolumn{2}{c}{Instacart}
& \multicolumn{2}{c}{Electronics}
& \multicolumn{2}{c}{Sports} \\
\cmidrule(lr){2-3}\cmidrule(lr){4-5}\cmidrule(lr){6-7}
%Metric
& N@10 & N@20
& N@10 & N@20
& N@10 & N@20 \\
\midrule

Ensemble: RC
& 0.0428 & 0.0633
& 0.0765 & 0.0874
& 0.0859 & 0.1009 \\
+ \modelname
& 0.0701 & 0.0825
& 0.1149 & 0.1227
& 0.1221 & 0.1286 \\
\midrule

Ensemble: LR
& 0.0401 & 0.0638
& 0.0862 & 0.0893
& 0.0950 & 0.1060 \\
+ \modelname
& \textbf{0.0796} & \textbf{0.0907}
& 0.1136 & 0.1200
& 0.1246 & 0.1297 \\
\midrule

Ensemble: MLP
& 0.0397 & 0.0635
& 0.0754 & 0.0861
& 0.1009 & 0.1056 \\
+ \modelname
& 0.0747 & 0.0858
& 0.1071 & 0.1166
& 0.1247 & 0.1328 \\
\midrule

Ensemble: Att
& 0.0473 & 0.0692
& 0.0787 & 0.0878
& 0.1000 & 0.1078 \\
+ \modelname
& 0.0765 & 0.0844
& \textbf{0.1180} & \textbf{0.1260}
& \textbf{0.1247} & \textbf{0.1326} \\
\bottomrule
\end{tabular}%
}
\vspace{-3mm}
\end{table}

% previous version
% \begin{table}[]\caption{Performance of different ensemble methods with and without all LLM reasoning modules from \modelname.}\label{tab:ensemble}
% \vspace{-3mm}
% \resizebox{0.475\textwidth}{!}{
% \begin{tabular}{l|cc|cc|cc}
% \hline
% \multicolumn{1}{c|}{Dataset} & \multicolumn{2}{c|}{Instacart}    & \multicolumn{2}{c|}{Electronics}       & \multicolumn{2}{c}{Sports}     \\ \hline
% Metric         & N@10 & N@20             & N@10             & N@20             & N@10            & N@20             \\ \hline
% Ensemble: RC & 0.0428 & 0.0633 & 0.0765 & 0.0874 & 0.0859	& 0.1009 \\
% + \modelname  & 0.0701 & 0.0825 & 0.1149 & 0.1227 & 0.1221 & 0.1286 \\ \hline
% Ensemble: LR  & 0.0401 & 0.0638 & 0.0862 & 0.0893 & 0.0950 & 0.1060 \\ 
% + \modelname & \textbf{0.0796} & \textbf{0.0907} & 0.1136 & 0.1200 & 0.1246 & 0.1297 \\ \hline
% Ensemble: MLP  & 0.0397 & 0.0635 & 0.0754 & 0.0861 & 0.1009 &0.1056 \\
% + \modelname & 0.0747 & 0.0858 & 0.1071 & 0.1166 & 0.1247 & 0.1328 \\ \hline
% Ensemble: Att  & 0.0473 & 0.0692 & 0.0787 & 0.0878 & 0.1000 & 0.1078 \\
% + \modelname & 0.0765 & 0.0844 & \textbf{0.1180} & \textbf{0.1260} & \textbf{0.1247} & \textbf{0.1326}
% \\ \hline
% \end{tabular}
% }
% \vspace{-3mm}
% \end{table}

\subsection{RQ4: Ablation Study}
% To quantify the contribution of each LLM-based module, we compare \modelname with different modules removed or replaced. 
% For fair comparison, we use the original ranking comparison as quantitative comparison module.
% The ranking comparison is used as quantitative comparison module.
\subsubsection{Tool Comparison and Dual S\&C Reranking}
\begin{figure}[]
    \begin{subfigure}{0.163\textwidth}
    \includegraphics[width=\textwidth]{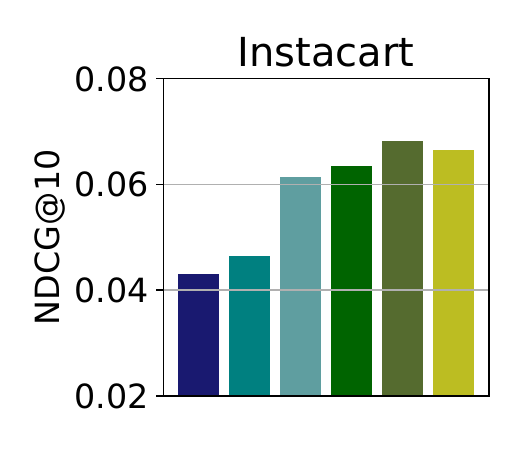}
    \end{subfigure}
    \hspace{-3mm}
    % \hfill
    \begin{subfigure}{0.163\textwidth}
    \includegraphics[width=\textwidth]{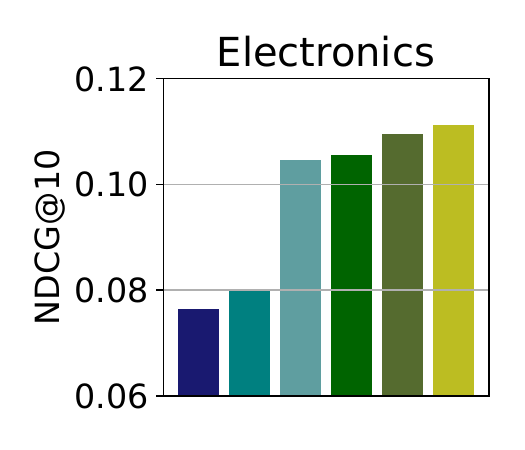}
    \end{subfigure}
    \hspace{-3mm}
    % \hfill
    \begin{subfigure}{0.163\textwidth}
    \includegraphics[width=\textwidth]{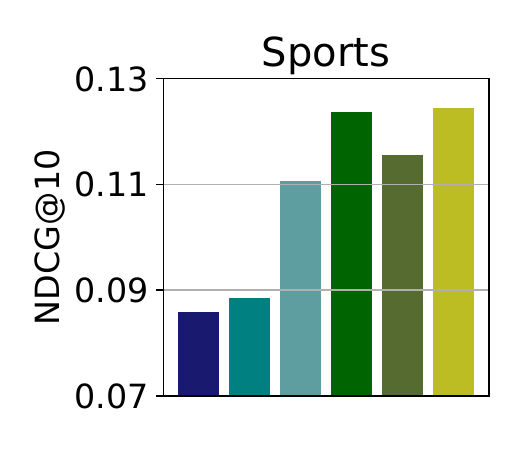}
    \end{subfigure}
    \begin{subfigure}{0.45\textwidth}
    \centering
   \vspace{-2mm}
    \includegraphics[width=\textwidth]{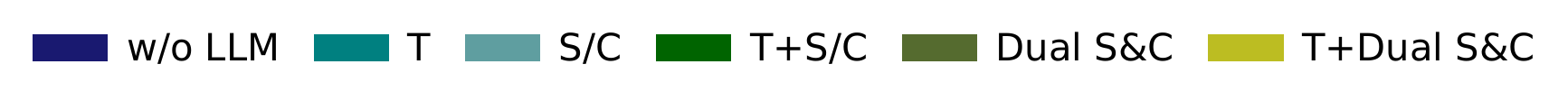}
    \end{subfigure}
    \vspace{-4mm}
        \caption{Ablation study on tool comparison or dual S\&C reranking modules. The bars of S/C denote reranking based on substitutes or complements according to intent memory.}
  \label{fig:ablation}
\end{figure}

We investigate the contribution of the proposed tool comparison and dual S\&C reranking modules, denoted as \textit{T} and \textit{Dual S\&C}, respetively. We also explore the effect of replacing dual S\&C by a mutually exclusive substitute or complement reranking based on intent memory of each user, denoted as \textit{S/C}. The general reranking module is excluded from this study. The results are shown in Fig.~\ref{fig:ablation}. We observe that tool comparison improves performance in most settings, while it may degrade performance on Instacart. In this dataset, the semantic differences between the grocery food lists recommended by different tools are less pronounced, providing weaker signals for LLM-based tool selection. Moreover, the dual S\&C reranking consistently outperforms reranking based on either substitutes or complements alone across all settings, highlighting the advantage of a more comprehensive reranking strategy.

\subsubsection{General Reranking}
The general reranking module in Sec.~\ref{sec:genrerank} refines recommendation lists based on summarized user preferences, independent of substitution and complementarity signals. In Table~\ref{tab:genrerank}, we compare the performance of \modelname with and without this module under different reranking settings. The results show that general reranking consistently improves performance in most cases, and all best results across the three datasets are achieved with this module enabled, confirming the importance of general preference signals in complementing intent-specific reasoning. Since the corresponding intent identification module in Eq.~\ref{eq:reglearn} accounts for one-third of all inference calls during agent optimization, it can be optionally deactivated to save computational resources in this flexible agent framework, without sacrificing the general preference signals. In such cases, $m_{reg}$ is set as a constant regularization coefficient shared across all users in the dataset.

\begin{table}
\vspace{-0.1cm}
\caption{Performance of different variants of \modelname with and without general reranking.}
\label{tab:genrerank}
\vspace{-3mm}
\resizebox{0.45\textwidth}{!}{%
\begin{tabular}{l cc @{\hspace{1.2em}} cc @{\hspace{1.2em}} cc}
\toprule
Dataset
& \multicolumn{2}{c}{Instacart}
& \multicolumn{2}{c}{Electronics}
& \multicolumn{2}{c}{Sports} \\
\cmidrule(lr){2-3}\cmidrule(lr){4-5}\cmidrule(lr){6-7}
Metric
& N@10 & N@20
& N@10 & N@20
& N@10 & N@20 \\
\midrule

S/C
& 0.0614 & 0.0740
& 0.1047 & 0.1076
& 0.1107 & 0.1174 \\
+ General
& 0.0676 & 0.0817
& 0.1026 & 0.1120
& 0.1109 & 0.1190 \\
\midrule

T+S/C
& 0.0634 & 0.0794
& 0.1054 & 0.1136
& 0.1238 & 0.1320 \\
+ General
& 0.0626 & 0.0800
& 0.1055 & 0.1149
& \textbf{0.1244} & \textbf{0.1323} \\
\midrule

Dual S\&C
& 0.0682 & 0.0805
& 0.1095 & 0.1174
& 0.1155 & 0.1203 \\
+ General
& 0.0684 & 0.0824
& 0.1050 & 0.1109
& 0.1190 & 0.1238 \\
\midrule

T+Dual S\&C
& 0.0666 & 0.0793
& 0.1112 & 0.1211
& 0.1240 & 0.1307 \\
+ General
& \textbf{0.0701} & \textbf{0.0825}
& \textbf{0.1149} & \textbf{0.1227}
& 0.1221 & 0.1286 \\
\bottomrule
\end{tabular}%
}
\vspace{-3mm}
\end{table}

% previous version
% \begin{table}[]\caption{Performance of different variants of \modelname with and without general reranking.}\label{tab:genrerank}
% \vspace{-3mm}
% \resizebox{0.475\textwidth}{!}{
% \begin{tabular}{l|cc|cc|cc}
% \hline
% \multicolumn{1}{c|}{Dataset} & \multicolumn{2}{c|}{Instacart}    & \multicolumn{2}{c|}{Electronics}       & \multicolumn{2}{c}{Sports}     \\ \hline
% Metric         & N@10 & N@20             & N@10             & N@20             & N@10            & N@20             \\ \hline
% S/C & 0.0614 & 0.0740 & 0.1047 & 0.1076 & 0.1107	& 0.1174 \\
% +General  & 0.0676  & 0.0817  & 0.1026 & 0.1120 & 0.1109 & 0.1190 \\ \hline
% T+S/C  & 0.0634 & 0.0794 & 0.1054 & 0.1136 & 0.1238 & 0.1320 \\ 
% +General & 0.0626 & 0.0800  & 0.1055 & 0.1149 & \textbf{0.1244} & \textbf{0.1323} \\ \hline
% Dual S\&C  & 0.0682 & 0.0805 & 0.1095 & 0.1174 & 0.1155 & 0.1203 \\
% + General & 0.0684  & 0.0824 & 0.1050 & 0.1109 & 0.1190 & 0.1238 \\ \hline
% T+Dual S\&C  & 0.0666  & 0.0793 & 0.1112 & 0.1211 & 0.1240 &0.1307 \\
% + General & \textbf{0.0701} & \textbf{0.0825} & \textbf{0.1149} & \textbf{0.1227} & 0.1221 & 0.1286
% \\\hline
% \end{tabular}
% }
% \vspace{-3mm}
% \end{table}

\section{Conclusion}
In this work, we propose a novel LLM-based agent framework, \modelname, for full-ranking recommendation. 
% It leverages the reasoning capabilities of LLMs to delegate the full-ranking task to multiple recommendation tools in a personalized manner. Furthermore, the commonsense world knowledge of the LLM is utilized to infer user intent regarding substitutes and complements, which is a challenging aspect for traditional models to capture effectively for recommendation enhancement. 
The commonsense world knowledge of the LLM is utilized to infer user intent regarding substitutes and complements.
% Extensive experiments are conducted with a proposed LLM-based evaluation metric VDCG to evaluate both semantic relevance and ordering correctness of recommendation results. 
A limitation of this work is that substitution and complementarity relationships are more prevalent in domains like groceries than in others such as movies or music. 
% Future work may explore knowledge sharing among user agents to better capture global data characteristics tailored to specific downstream scenarios.

\section*{Acknowledgements}
This work is supported in part by NSF under grants III-2106758, and POSE-2346158.

\bibliographystyle{ACM-Reference-Format}
\balance
\bibliography{AgentDR}

@inproceedings{YangLYLW0Y25,
  author       = {Mingdai Yang and
                  Zhiwei Liu and
                  Liangwei Yang and
                  Xiaolong Liu and
                  Chen Wang and
                  Hao Peng and
                  Philip S. Yu},
 _editor       = {Guodong Long and
                  Michale Blumestein and
                  Yi Chang and
                  Liane Lewin{-}Eytan and
                  Zi Helen Huang and
                  Elad Yom{-}Tov},
  title        = {Training Large Recommendation Models via Graph-Language Tokens Alignment},
  booktitle    = {Companion Proceedings of the {ACM} on Web Conference 2025, {WWW} 2025,
                  Sydney, NSW, Australia, 28 April 2025 - 2 May 2025},
  pages        = {1470--1474},
  publisher    = {{ACM}},
  year         = {2025},
  timestamp    = {Sun, 02 Nov 2025 21:27:17 +0100},
  bibsource    = {dblp computer science bibliography, https://dblp.org}
}

@inproceedings{simplex,
  author       = {Kelong Mao and
                  Jieming Zhu and
                  Jinpeng Wang and
                  Quanyu Dai and
                  Zhenhua Dong and
                  Xi Xiao and
                  Xiuqiang He},
 _editor       = {Gianluca Demartini and
                  Guido Zuccon and
                  J. Shane Culpepper and
                  Zi Huang and
                  Hanghang Tong},
  title        = {SimpleX: {A} Simple and Strong Baseline for Collaborative Filtering},
  booktitle    = {{CIKM} '21: The 30th {ACM} International Conference on Information
                  and Knowledge Management, Virtual Event, Queensland, Australia, November
                  1 - 5, 2021},
  pages        = {1243--1252},
  publisher    = {{ACM}},
  year         = {2021},
  timestamp    = {Sun, 19 Jan 2025 13:12:43 +0100},
  bibsource    = {dblp computer science bibliography, https://dblp.org}
}

@inproceedings{sasrec,
  author       = {Wang{-}Cheng Kang and
                  Julian J. McAuley},
  title        = {Self-Attentive Sequential Recommendation},
  booktitle    = {{IEEE} International Conference on Data Mining, {ICDM} 2018, Singapore,
                  November 17-20, 2018},
  pages        = {197--206},
  publisher    = {{IEEE} Computer Society},
  year         = {2018},
  bibsource    = {dblp computer science bibliography, https://dblp.org}
}

@inproceedings{recmind,
    title = "{R}ec{M}ind: Large Language Model Powered Agent For Recommendation",
    author = "Wang, Yancheng  and
      Jiang, Ziyan  and
      Chen, Zheng  and
      Yang, Fan  and
      Zhou, Yingxue  and
      Cho, Eunah  and
      Fan, Xing  and
      Lu, Yanbin  and
      Huang, Xiaojiang  and
      Yang, Yingzhen",
    _editor = "Duh, Kevin  and
      Gomez, Helena  and
      Bethard, Steven",
    booktitle = "Findings of the Association for Computational Linguistics: NAACL 2024",
    month = jun,
    year = "2024",
    address = "Mexico City, Mexico",
    publisher = "Association for Computational Linguistics",
    pages = "4351--4364",
}

@article{recagent,
author = {Wang, Lei and Zhang, Jingsen and Yang, Hao and Chen, Zhi-Yuan and Tang, Jiakai and Zhang, Zeyu and Chen, Xu and Lin, Yankai and Sun, Hao and Song, Ruihua and Zhao, Xin and Xu, Jun and Dou, Zhicheng and Wang, Jun and Wen, Ji-Rong},
title = {User Behavior Simulation with Large Language Model-based Agents},
year = {2025},
issue_date = {March 2025},
publisher = {Association for Computing Machinery},
address = {New York, NY, USA},
volume = {43},
number = {2},
issn = {1046-8188},
journal = {ACM Trans. Inf. Syst.},
month = jan,
articleno = {55},
numpages = {37}
}

@inproceedings{ZhongGGYW24,
  author       = {Wanjun Zhong and
                  Lianghong Guo and
                  Qiqi Gao and
                  He Ye and
                  Yanlin Wang},
 _editor       = {Michael J. Wooldridge and
                  Jennifer G. Dy and
                  Sriraam Natarajan},
  title        = {MemoryBank: Enhancing Large Language Models with Long-Term Memory},
  booktitle    = {Thirty-Eighth {AAAI} Conference on Artificial Intelligence, {AAAI}
                  2024, Thirty-Sixth Conference on Innovative Applications of Artificial
                  Intelligence, {IAAI} 2024, Fourteenth Symposium on Educational Advances
                  in Artificial Intelligence, {EAAI} 2014, February 20-27, 2024, Vancouver,
                  Canada},
  pages        = {19724--19731},
  publisher    = {{AAAI} Press},
  year         = {2024},
  timestamp    = {Tue, 04 Mar 2025 08:09:48 +0100},
  bibsource    = {dblp computer science bibliography, https://dblp.org}
}

@inproceedings{ShinnCGNY23,
  author       = {Noah Shinn and
                  Federico Cassano and
                  Ashwin Gopinath and
                  Karthik Narasimhan and
                  Shunyu Yao},
  _editor       = {Alice Oh and
                  Tristan Naumann and
                  Amir Globerson and
                  Kate Saenko and
                  Moritz Hardt and
                  Sergey Levine},
  title        = {Reflexion: language agents with verbal reinforcement learning},
  booktitle    = {Advances in Neural Information Processing Systems 36: Annual Conference
                  on Neural Information Processing Systems 2023, NeurIPS 2023, New Orleans,
                  LA, USA, December 10 - 16, 2023},
  year         = {2023},
  timestamp    = {Fri, 01 Mar 2024 16:26:19 +0100},
  bibsource    = {dblp computer science bibliography, https://dblp.org}
}

@inproceedings{gao2025,
    title = "An Efficient Context-Dependent Memory Framework for {LLM}-Centric Agents",
    author = "Gao, Pengyu  and
      Zhao, Jinming  and
      Chen, Xinyue  and
      Yilin, Long",
   _editor = "Chen, Weizhu  and
      Yang, Yi  and
      Kachuee, Mohammad  and
      Fu, Xue-Yong",
    booktitle = "Proceedings of the 2025 Conference of the Nations of the Americas Chapter of the Association for Computational Linguistics: Human Language Technologies (Volume 3: Industry Track)",
    month = apr,
    year = "2025",
    address = "Albuquerque, New Mexico",
    publisher = "Association for Computational Linguistics",
    pages = "1055--1069",
    ISBN = "979-8-89176-194-0",
}

@inproceedings{agentcf,
  author       = {Junjie Zhang and
                  Yupeng Hou and
                  Ruobing Xie and
                  Wenqi Sun and
                  Julian J. McAuley and
                  Wayne Xin Zhao and
                  Leyu Lin and
                  Ji{-}Rong Wen},
 _editor       = {Tat{-}Seng Chua and
                  Chong{-}Wah Ngo and
                  Ravi Kumar and
                  Hady W. Lauw and
                  Roy Ka{-}Wei Lee},
  title        = {AgentCF: Collaborative Learning with Autonomous Language Agents for
                  Recommender Systems},
  booktitle    = {Proceedings of the {ACM} on Web Conference 2024, {WWW} 2024, Singapore,
                  May 13-17, 2024},
  pages        = {3679--3689},
  publisher    = {{ACM}},
  year         = {2024},
  timestamp    = {Sun, 19 Jan 2025 13:10:29 +0100},
  bibsource    = {dblp computer science bibliography, https://dblp.org}
}

@inproceedings{agent4rec,
  author       = {An Zhang and
                  Yuxin Chen and
                  Leheng Sheng and
                  Xiang Wang and
                  Tat{-}Seng Chua},
 _editor       = {Grace Hui Yang and
                  Hongning Wang and
                  Sam Han and
                  Claudia Hauff and
                  Guido Zuccon and
                  Yi Zhang},
  title        = {On Generative Agents in Recommendation},
  booktitle    = {Proceedings of the 47th International {ACM} {SIGIR} Conference on
                  Research and Development in Information Retrieval, {SIGIR} 2024, Washington
                  DC, USA, July 14-18, 2024},
  pages        = {1807--1817},
  publisher    = {{ACM}},
  year         = {2024},
  timestamp    = {Sun, 19 Jan 2025 13:11:20 +0100},
  bibsource    = {dblp computer science bibliography, https://dblp.org}
}

@article{hallu1,
author = {Yu, Chung-En and Jalaian, Brian and Bastian, Nathaniel},
year = {2024},
month = {11},
pages = {110-113},
title = {Mitigating Large Vision-Language Model Hallucination at Post-hoc via Multi-agent System},
volume = {4},
journal = {Proceedings of the AAAI Symposium Series},
}

@inproceedings{hallu2,
author = {Shen, Yedan and Wu, Kaixin and Ding, Yuechen and Wen, Jingyuan and Liu, Hong and Zhong, Mingjie and Lin, Zhouhan and Xu, Jia and Mo, Linjian},
title = {Alleviating LLM-based Generative Retrieval Hallucination in Alipay Search},
year = {2025},
isbn = {9798400715921},
publisher = {Association for Computing Machinery},
address = {New York, NY, USA},
booktitle = {Proceedings of the 48th International ACM SIGIR Conference on Research and Development in Information Retrieval},
pages = {4294–4298},
numpages = {5},
location = {Padua, Italy},
series = {SIGIR '25}
}

@inproceedings{coral,
  author       = {Junda Wu and
                  Cheng{-}Chun Chang and
                  Tong Yu and
                  Zhankui He and
                  Jianing Wang and
                  Yupeng Hou and
                  Julian J. McAuley},
 _editor       = {Ricardo Baeza{-}Yates and
                  Francesco Bonchi},
  title        = {CoRAL: Collaborative Retrieval-Augmented Large Language Models Improve
                  Long-tail Recommendation},
  booktitle    = {Proceedings of the 30th {ACM} {SIGKDD} Conference on Knowledge Discovery
                  and Data Mining, {KDD} 2024, Barcelona, Spain, August 25-29, 2024},
  pages        = {3391--3401},
  publisher    = {{ACM}},
  year         = {2024},
  timestamp    = {Mon, 23 Jun 2025 08:08:08 +0200},
  bibsource    = {dblp computer science bibliography, https://dblp.org}
}

@inproceedings{kgrag,
    title = "Knowledge Graph Retrieval-Augmented Generation for {LLM}-based Recommendation",
    author = "Wang, Shijie  and
      Fan, Wenqi  and
      Feng, Yue  and
      Shanru, Lin  and
      Ma, Xinyu  and
      Wang, Shuaiqiang  and
      Yin, Dawei",
    _editor = "Che, Wanxiang  and
      Nabende, Joyce  and
      Shutova, Ekaterina  and
      Pilehvar, Mohammad Taher",
    booktitle = "Proceedings of the 63rd Annual Meeting of the Association for Computational Linguistics (Volume 1: Long Papers)",
    month = jul,
    year = "2025",
    address = "Vienna, Austria",
    publisher = "Association for Computational Linguistics",
    pages = "27152--27168",
    ISBN = "979-8-89176-251-0"
}

@inproceedings{HuangBYJ0SWKY25,
  author       = {Feiran Huang and
                  Yuanchen Bei and
                  Zhenghang Yang and
                  Junyi Jiang and
                  Hao Chen and
                  Qijie Shen and
                  Senzhang Wang and
                  Fakhri Karray and
                  Philip S. Yu},
 _editor       = {Wolfgang Nejdl and
                  S{\"{o}}ren Auer and
                  Meeyoung Cha and
                  Marie{-}Francine Moens and
                  Marc Najork},
  title        = {Large Language Model Simulator for Cold-Start Recommendation},
  booktitle    = {Proceedings of the Eighteenth {ACM} International Conference on Web
                  Search and Data Mining, {WSDM} 2025, Hannover, Germany, March 10-14,
                  2025},
  pages        = {261--270},
  publisher    = {{ACM}},
  year         = {2025},
  timestamp    = {Fri, 07 Mar 2025 18:29:44 +0100},
  bibsource    = {dblp computer science bibliography, https://dblp.org}
}

@inproceedings{llmrank,
  author       = {Yupeng Hou and
                  Junjie Zhang and
                  Zihan Lin and
                  Hongyu Lu and
                  Ruobing Xie and
                  Julian J. McAuley and
                  Wayne Xin Zhao},
 _editor       = {Nazli Goharian and
                  Nicola Tonellotto and
                  Yulan He and
                  Aldo Lipani and
                  Graham McDonald and
                  Craig Macdonald and
                  Iadh Ounis},
  title        = {Large Language Models are Zero-Shot Rankers for Recommender Systems},
  booktitle    = {Advances in Information Retrieval - 46th European Conference on Information
                  Retrieval, {ECIR} 2024, Glasgow, UK, March 24-28, 2024, Proceedings,
                  Part {II}},
  series       = {Lecture Notes in Computer Science},
  volume       = {14609},
  pages        = {364--381},
  publisher    = {Springer},
  year         = {2024},
  timestamp    = {Tue, 27 Aug 2024 07:06:51 +0200},
  bibsource    = {dblp computer science bibliography, https://dblp.org}
}

@inproceedings{rlmrec,
  author       = {Xubin Ren and
                  Wei Wei and
                  Lianghao Xia and
                  Lixin Su and
                  Suqi Cheng and
                  Junfeng Wang and
                  Dawei Yin and
                  Chao Huang},
  title        = {Representation Learning with Large Language Models for Recommendation},
  booktitle    = {Proceedings of the {ACM} on Web Conference},
  pages        = {3464--3475},
  publisher    = {{ACM}},
  year         = {2024},
}

@inproceedings{p5,
  author       = {Shijie Geng and
                  Shuchang Liu and
                  Zuohui Fu and
                  Yingqiang Ge and
                  Yongfeng Zhang},
   title        = {Recommendation as Language Processing {(RLP):} {A} Unified Pretrain,
                  Personalized Prompt {\&} Predict Paradigm {(P5)}},
  booktitle    = {RecSys '22: Sixteenth {ACM} Conference on Recommender Systems},
  pages        = {299--315},
  publisher    = {{ACM}},
  year         = {2022},
}

@inproceedings{ihp,
  author       = {Mingdai Yang and
                  Zhiwei Liu and
                  Liangwei Yang and
                  Xiaolong Liu and
                  Chen Wang and
                  Hao Peng and
                  Philip S. Yu},
  title        = {Instruction-based Hypergraph Pretraining},
  booktitle    = {Proceedings of the 47th International {ACM} {SIGIR} Conference on
                  Research and Development in Information Retrieval},
  pages        = {501--511},
  publisher    = {{ACM}},
  year         = {2024},
  bibsource    = {dblp computer science bibliography, https://dblp.org}
}

@article{bm25,
  author       = {Stephen E. Robertson and
                  Hugo Zaragoza},
  title        = {The Probabilistic Relevance Framework: {BM25} and Beyond},
  journal      = {Found. Trends Inf. Retr.},
  volume       = {3},
  number       = {4},
  pages        = {333--389},
  year         = {2009},
  bibsource    = {dblp computer science bibliography, https://dblp.org}
}

@article{enmf,
  author       = {Chong Chen and
                  Min Zhang and
                  Yongfeng Zhang and
                  Yiqun Liu and
                  Shaoping Ma},
  title        = {Efficient Neural Matrix Factorization without Sampling for Recommendation},
  journal      = {{ACM} Trans. Inf. Syst.},
  volume       = {38},
  number       = {2},
  pages        = {14:1--14:28},
  year         = {2020},
  bibsource    = {dblp computer science bibliography, https://dblp.org}
}

@inproceedings{diffrec,
  author       = {Wenjie Wang and
                  Yiyan Xu and
                  Fuli Feng and
                  Xinyu Lin and
                  Xiangnan He and
                  Tat{-}Seng Chua},
 _editor       = {Hsin{-}Hsi Chen and
                  Wei{-}Jou (Edward) Duh and
                  Hen{-}Hsen Huang and
                  Makoto P. Kato and
                  Josiane Mothe and
                  Barbara Poblete},
  title        = {Diffusion Recommender Model},
  booktitle    = {Proceedings of the 46th International {ACM} {SIGIR} Conference on
                  Research and Development in Information Retrieval, {SIGIR} 2023, Taipei,
                  Taiwan, July 23-27, 2023},
  pages        = {832--841},
  publisher    = {{ACM}},
  year         = {2023},
  timestamp    = {Tue, 18 Feb 2025 15:22:19 +0100},
  bibsource    = {dblp computer science bibliography, https://dblp.org}
}

@inproceedings{fearec,
  author       = {Xinyu Du and
                  Huanhuan Yuan and
                  Pengpeng Zhao and
                  Jianfeng Qu and
                  Fuzhen Zhuang and
                  Guanfeng Liu and
                  Yanchi Liu and
                  Victor S. Sheng},
 _editor       = {Hsin{-}Hsi Chen and
                  Wei{-}Jou (Edward) Duh and
                  Hen{-}Hsen Huang and
                  Makoto P. Kato and
                  Josiane Mothe and
                  Barbara Poblete},
  title        = {Frequency Enhanced Hybrid Attention Network for Sequential Recommendation},
  booktitle    = {Proceedings of the 46th International {ACM} {SIGIR} Conference on
                  Research and Development in Information Retrieval, {SIGIR} 2023, Taipei,
                  Taiwan, July 23-27, 2023},
  pages        = {78--88},
  publisher    = {{ACM}},
  year         = {2023},
  timestamp    = {Thu, 30 Jan 2025 15:13:33 +0100},
  bibsource    = {dblp computer science bibliography, https://dblp.org}
}

@inproceedings{lacklabel21,
author = {Bonab, Hamed and Aliannejadi, Mohammad and Vardasbi, Ali and Kanoulas, Evangelos and Allan, James},
title = {Cross-Market Product Recommendation},
year = {2021},
isbn = {9781450384469},
publisher = {Association for Computing Machinery},
address = {New York, NY, USA},
booktitle = {Proceedings of the 30th ACM International Conference on Information \& Knowledge Management},
pages = {110–119},
numpages = {10},
location = {Virtual Event, Queensland, Australia},
series = {CIKM '21}
}

@inproceedings{sclabel20chen,
author = {Chen, Tong and Yin, Hongzhi and Ye, Guanhua and Huang, Zi and Wang, Yang and Wang, Meng},
title = {Try This Instead: Personalized and Interpretable Substitute Recommendation},
year = {2020},
isbn = {9781450380164},
publisher = {Association for Computing Machinery},
address = {New York, NY, USA},
booktitle = {Proceedings of the 43rd International ACM SIGIR Conference on Research and Development in Information Retrieval},
pages = {891–900},
numpages = {10},
location = {Virtual Event, China},
series = {SIGIR '20}
}

@inproceedings{sclabel20liu,
author = {Liu, Yiding and Gu, Yulong and Ding, Zhuoye and Gao, Junchao and Guo, Ziyi and Bao, Yongjun and Yan, Weipeng},
title = {Decoupled Graph Convolution Network for Inferring Substitutable and Complementary Items},
year = {2020},
isbn = {9781450368599},
publisher = {Association for Computing Machinery},
address = {New York, NY, USA},
booktitle = {Proceedings of the 29th ACM International Conference on Information \& Knowledge Management},
pages = {2621–2628},
numpages = {8},
location = {Virtual Event, Ireland},
series = {CIKM '20}
}

@inproceedings{sclabel20xu,
author = {Xu, Da and Ruan, Chuanwei and Korpeoglu, Evren and Kumar, Sushant and Achan, Kannan},
title = {Product Knowledge Graph Embedding for E-commerce},
year = {2020},
isbn = {9781450368223},
publisher = {Association for Computing Machinery},
address = {New York, NY, USA},
booktitle = {Proceedings of the 13th International Conference on Web Search and Data Mining},
pages = {672–680},
numpages = {9},
location = {Houston, TX, USA},
series = {WSDM '20}
}

@inproceedings{lacklabel19,
    title = "Justifying Recommendations using Distantly-Labeled Reviews and Fine-Grained Aspects",
    author = "Ni, Jianmo  and
      Li, Jiacheng  and
      McAuley, Julian",
   _editor = "Inui, Kentaro  and
      Jiang, Jing  and
      Ng, Vincent  and
      Wan, Xiaojun",
    booktitle = "Proceedings of the 2019 Conference on Empirical Methods in Natural Language Processing and the 9th International Joint Conference on Natural Language Processing (EMNLP-IJCNLP)",
    month = nov,
    year = "2019",
    address = "Hong Kong, China",
    publisher = "Association for Computational Linguistics",
    pages = "188--197",
}

@inproceedings{xmarket,
  author       = {Hamed R. Bonab and
                  Mohammad Aliannejadi and
                  Ali Vardasbi and
                  Evangelos Kanoulas and
                  James Allan},
 _editor       = {Gianluca Demartini and
                  Guido Zuccon and
                  J. Shane Culpepper and
                  Zi Huang and
                  Hanghang Tong},
  title        = {Cross-Market Product Recommendation},
  booktitle    = {{CIKM} '21: The 30th {ACM} International Conference on Information
                  and Knowledge Management, Virtual Event, Queensland, Australia, November
                  1 - 5, 2021},
  pages        = {110--119},
  publisher    = {{ACM}},
  year         = {2021},
  timestamp    = {Sat, 30 Sep 2023 09:36:59 +0200},
  bibsource    = {dblp computer science bibliography, https://dblp.org}
}

@inproceedings{dcg,
author = {Burges, Chris and Shaked, Tal and Renshaw, Erin and Lazier, Ari and Deeds, Matt and Hamilton, Nicole and Hullender, Greg},
title = {Learning to rank using gradient descent},
year = {2005},
isbn = {1595931805},
publisher = {Association for Computing Machinery},
address = {New York, NY, USA},
booktitle = {Proceedings of the 22nd International Conference on Machine Learning},
pages = {89–96},
numpages = {8},
location = {Bonn, Germany},
series = {ICML '05}
}

@article{hallulength1,
  author       = {Lei Huang and
                  Weijiang Yu and
                  Weitao Ma and
                  Weihong Zhong and
                  Zhangyin Feng and
                  Haotian Wang and
                  Qianglong Chen and
                  Weihua Peng and
                  Xiaocheng Feng and
                  Bing Qin and
                  Ting Liu},
  title        = {A Survey on Hallucination in Large Language Models: Principles, Taxonomy,
                  Challenges, and Open Questions},
  journal      = {{ACM} Trans. Inf. Syst.},
  volume       = {43},
  number       = {2},
  pages        = {42:1--42:55},
  year         = {2025},
  bibsource    = {dblp computer science bibliography, https://dblp.org}
}

@article{m6rec,
  author       = {Zeyu Cui and
                  Jianxin Ma and
                  Chang Zhou and
                  Jingren Zhou and
                  Hongxia Yang},
  title        = {M6-Rec: Generative Pretrained Language Models are Open-Ended Recommender
                  Systems},
  journal      = {CoRR},
  volume       = {abs/2205.08084},
  year         = {2022},
  eprinttype    = {arXiv},
  eprint       = {2205.08084},
  bibsource    = {dblp computer science bibliography, https://dblp.org}
}

@inproceedings{iagent,
    title = "i{A}gent: {LLM} Agent as a Shield between User and Recommender Systems",
    author = "Xu, Wujiang  and
      Shi, Yunxiao  and
      Liang, Zujie  and
      Ning, Xuying  and
      Mei, Kai  and
      Wang, Kun  and
      Zhu, Xi  and
      Xu, Min  and
      Zhang, Yongfeng",
   _editor = "Che, Wanxiang  and
      Nabende, Joyce  and
      Shutova, Ekaterina  and
      Pilehvar, Mohammad Taher",
    booktitle = "Findings of the Association for Computational Linguistics: ACL 2025",
    month = jul,
    year = "2025",
    address = "Vienna, Austria",
    publisher = "Association for Computational Linguistics",
    pages = "18056--18084",
    ISBN = "979-8-89176-256-5"
}

@inproceedings{llmlabel25Hosseini,
author = {Hosseini, Kasra and Kober, Thomas and Krapac, Josip and Vollgraf, Roland and Cheng, Weiwei and Peleteiro Ramallo, Ana},
title = {Retrieve, Annotate, Evaluate, Repeat: Leveraging Multimodal LLMs for Large-Scale Product Retrieval Evaluation},
year = {2025},
isbn = {978-3-031-88707-9},
publisher = {Springer-Verlag},
address = {Berlin, Heidelberg},
booktitle = {Advances in Information Retrieval: 47th European Conference on Information Retrieval, ECIR 2025, Lucca, Italy, April 6–10, 2025, Proceedings, Part I},
pages = {149–163},
numpages = {15},
location = {Lucca, Italy}
}

@inproceedings{llmlabel23papso,
author = {Papso, Rastislav},
title = {Complementary Product Recommendation for Long-tail Products},
year = {2023},
isbn = {9798400702419},
publisher = {Association for Computing Machinery},
address = {New York, NY, USA},
booktitle = {Proceedings of the 17th ACM Conference on Recommender Systems},
pages = {1305–1311},
numpages = {7},
location = {Singapore, Singapore},
series = {RecSys '23}
}

@inproceedings{notrealsc24kai,
author = {Sugahara, Kai and Yamasaki, Chihiro and Okamoto, Kazushi},
title = {Is It Really Complementary? Revisiting Behavior-based Labels for Complementary Recommendation},
year = {2024},
isbn = {9798400705052},
publisher = {Association for Computing Machinery},
address = {New York, NY, USA},
booktitle = {Proceedings of the 18th ACM Conference on Recommender Systems},
pages = {1091–1095},
numpages = {5},
location = {Bari, Italy},
series = {RecSys '24}
}

@inproceedings{notrealsc23ye,
author = {Ye, Wenting and Yang, Hongfei and Zhao, Shuai and Fang, Haoyang and Shi, Xingjian and Neppalli, Naveen},
title = {A Transformer-Based Substitute Recommendation Model Incorporating Weakly Supervised Customer Behavior Data},
year = {2023},
isbn = {9781450394086},
publisher = {Association for Computing Machinery},
address = {New York, NY, USA},
booktitle = {Proceedings of the 46th International ACM SIGIR Conference on Research and Development in Information Retrieval},
pages = {3325–3329},
numpages = {5},
location = {Taipei, Taiwan},
series = {SIGIR '23}
}

@inproceedings{notrealsc22tong,
  author       = {Tong Jian and
                  Fan Yang and
                  Zhen Zuo and
                  Wenbo Wang and
                  Michinari Momma and
                  Tong Zhao and
                  Chaosheng Dong and
                  Yan Gao and
                  Yi Sun},
 _editor       = {Fr{\'{e}}d{\'{e}}rique Laforest and
                  Rapha{\"{e}}l Troncy and
                  Elena Simperl and
                  Deepak Agarwal and
                  Aristides Gionis and
                  Ivan Herman and
                  Lionel M{\'{e}}dini},
  title        = {Multi-task {GNN} for Substitute Identification},
  booktitle    = {Companion of The Web Conference 2022, Virtual Event / Lyon, France,
                  April 25 - 29, 2022},
  pages        = {228--231},
  publisher    = {{ACM}},
  year         = {2022},
  timestamp    = {Sat, 30 Sep 2023 09:59:31 +0200},
  bibsource    = {dblp computer science bibliography, https://dblp.org}
}

@article{subcom22,
  author       = {Wei Zhang and
                  Zeyuan Chen and
                  Hongyuan Zha and
                  Jianyong Wang},
  title        = {Learning from Substitutable and Complementary Relations for Graph-based
                  Sequential Product Recommendation},
  journal      = {{ACM} Trans. Inf. Syst.},
  volume       = {40},
  number       = {2},
  pages        = {26:1--26:28},
  year         = {2022},
  bibsource    = {dblp computer science bibliography, https://dblp.org}
}

@article{subcom23,
author = {Wu, Huizi and Geng, Cong and Fang, Hui},
title = {Session-based recommendation by exploiting substitutable and complementary relationships from multi-behavior data},
year = {2023},
issue_date = {May 2024},
publisher = {Kluwer Academic Publishers},
address = {USA},
volume = {38},
number = {3},
issn = {1384-5810},
journal = {Data Min. Knowl. Discov.},
month = dec,
pages = {1193–1221},
numpages = {29}
}

@inproceedings{LightGCN,
  author       = {Xiangnan He and
                  Kuan Deng and
                  Xiang Wang and
                  Yan Li and
                  Yong{-}Dong Zhang and
                  Meng Wang},
  title        = {LightGCN: Simplifying and Powering Graph Convolution Network for Recommendation},
  booktitle    = {Proceedings of the 43rd International {ACM} {SIGIR} conference on
                  research and development in Information Retrieval},
  pages        = {639--648},
  publisher    = {{ACM}},
  year         = {2020},
}

@inproceedings{LiDWCF24,
  author       = {Li Tao Li and
                  Steven H. H. Ding and
                  Andrew Walenstein and
                  Philippe Charland and
                  Benjamin C. M. Fung},
 _editor       = {Edoardo Serra and
                  Francesca Spezzano},
  title        = {Dynamic Neural Control Flow Execution: an Agent-Based Deep Equilibrium
                  Approach for Binary Vulnerability Detection},
  booktitle    = {Proceedings of the 33rd {ACM} International Conference on Information
                  and Knowledge Management, {CIKM} 2024, Boise, ID, USA, October 21-25,
                  2024},
  pages        = {1215--1225},
  publisher    = {{ACM}},
  year         = {2024},
  timestamp    = {Sun, 19 Jan 2025 13:12:29 +0100},
  bibsource    = {dblp computer science bibliography, https://dblp.org}
}

@inproceedings{ShiJQY24,
  author       = {Yuchen Shi and
                  Guochao Jiang and
                  Tian Qiu and
                  Deqing Yang},
 _editor       = {Edoardo Serra and
                  Francesca Spezzano},
  title        = {AgentRE: An Agent-Based Framework for Navigating Complex Information
                  Landscapes in Relation Extraction},
  booktitle    = {Proceedings of the 33rd {ACM} International Conference on Information
                  and Knowledge Management, {CIKM} 2024, Boise, ID, USA, October 21-25,
                  2024},
  pages        = {2045--2055},
  publisher    = {{ACM}},
  year         = {2024},
  timestamp    = {Sun, 19 Jan 2025 13:12:29 +0100},
  bibsource    = {dblp computer science bibliography, https://dblp.org}
}

@inproceedings{ZhaoTZC024,
  author       = {Runhao Zhao and
                  Jiuyang Tang and
                  Weixin Zeng and
                  Ziyang Chen and
                  Xiang Zhao},
 _editor       = {Edoardo Serra and
                  Francesca Spezzano},
  title        = {Zero-shot Knowledge Graph Question Generation via Multi-agent LLMs
                  and Small Models Synthesis},
  booktitle    = {Proceedings of the 33rd {ACM} International Conference on Information
                  and Knowledge Management, {CIKM} 2024, Boise, ID, USA, October 21-25,
                  2024},
  pages        = {3341--3351},
  publisher    = {{ACM}},
  year         = {2024},
  timestamp    = {Sun, 19 Jan 2025 13:12:33 +0100},
  bibsource    = {dblp computer science bibliography, https://dblp.org}
}

@article{interecagent,
author = {Huang, Xu and Lian, Jianxun and Lei, Yuxuan and Yao, Jing and Lian, Defu and Xie, Xing},
title = {Recommender AI Agent: Integrating Large Language Models for Interactive Recommendations},
year = {2025},
publisher = {Association for Computing Machinery},
address = {New York, NY, USA},
issn = {1046-8188},
journal = {ACM Trans. Inf. Syst.},
month = apr
}

@inproceedings{NEURIPS2024_2db8ce96,
 author = {Wu, Shirley and Zhao, Shiyu and Huang, Qian and Huang, Kexin and Yasunaga, Michihiro and Cao, Kaidi and Ioannidis, Vassilis N. and Subbian, Karthik and Leskovec, Jure and Zou, James},
 booktitle = {Advances in Neural Information Processing Systems},
_editor = {A. Globerson and L. Mackey and D. Belgrave and A. Fan and U. Paquet and J. Tomczak and C. Zhang},
 pages = {25981--26010},
 publisher = {Curran Associates, Inc.},
 title = {AvaTaR: Optimizing LLM Agents for Tool Usage via Contrastive Reasoning},
 volume = {37},
 year = {2024}
}

@inproceedings{0007XCW25,
  author       = {Hang Li and
                  Tianlong Xu and
                  Ethan Chang and
                  Qingsong Wen},
 _editor       = {Toby Walsh and
                  Julie Shah and
                  Zico Kolter},
  title        = {Knowledge Tagging with Large Language Model Based Multi-Agent System},
  booktitle    = {AAAI-25, Sponsored by the Association for the Advancement of Artificial
                  Intelligence, February 25 - March 4, 2025, Philadelphia, PA, {USA}},
  pages        = {28775--28782},
  publisher    = {{AAAI} Press},
  year         = {2025},
  timestamp    = {Thu, 17 Apr 2025 17:08:58 +0200},
  bibsource    = {dblp computer science bibliography, https://dblp.org}
}

@inproceedings{ZhuWGHL24,
  author       = {Yaochen Zhu and
                  Liang Wu and
                  Qi Guo and
                  Liangjie Hong and
                  Jundong Li},
 _editor       = {Tat{-}Seng Chua and
                  Chong{-}Wah Ngo and
                  Ravi Kumar and
                  Hady W. Lauw and
                  Roy Ka{-}Wei Lee},
  title        = {Collaborative Large Language Model for Recommender Systems},
  booktitle    = {Proceedings of the {ACM} on Web Conference 2024, {WWW} 2024, Singapore,
                  May 13-17, 2024},
  pages        = {3162--3172},
  publisher    = {{ACM}},
  year         = {2024},
  timestamp    = {Sun, 19 Jan 2025 13:10:18 +0100},
  bibsource    = {dblp computer science bibliography, https://dblp.org}
}

@inproceedings{LiaoL0WYW024,
  author       = {Jiayi Liao and
                  Sihang Li and
                  Zhengyi Yang and
                  Jiancan Wu and
                  Yancheng Yuan and
                  Xiang Wang and
                  Xiangnan He},
 _editor       = {Grace Hui Yang and
                  Hongning Wang and
                  Sam Han and
                  Claudia Hauff and
                  Guido Zuccon and
                  Yi Zhang},
  title        = {LLaRA: Large Language-Recommendation Assistant},
  booktitle    = {Proceedings of the 47th International {ACM} {SIGIR} Conference on
                  Research and Development in Information Retrieval, {SIGIR} 2024, Washington
                  DC, USA, July 14-18, 2024},
  pages        = {1785--1795},
  publisher    = {{ACM}},
  year         = {2024},
  timestamp    = {Fri, 07 Feb 2025 12:35:35 +0100},
  bibsource    = {dblp computer science bibliography, https://dblp.org}
}

@inproceedings{llmrec,
  author       = {Wei Wei and
                  Xubin Ren and
                  Jiabin Tang and
                  Qinyong Wang and
                  Lixin Su and
                  Suqi Cheng and
                  Junfeng Wang and
                  Dawei Yin and
                  Chao Huang},
  title        = {LLMRec: Large Language Models with Graph Augmentation for Recommendation},
  booktitle    = {Proceedings of the 17th {ACM} International Conference on Web Search
                  and Data Mining},
  pages        = {806--815},
  publisher    = {{ACM}},
  year         = {2024},
}

@inproceedings{pcl,
  author       = {Mingdai Yang and
                  Fan Yang and
                  Yanhui Guo and
                  Shaoyuan Xu and
                  Tianchen Zhou and
                  Yetian Chen and
                  Simone Shao and
                  Jia Liu and
                  Yan Gao},
 _editor       = {Guodong Long and
                  Michale Blumestein and
                  Yi Chang and
                  Liane Lewin{-}Eytan and
                  Zi Helen Huang and
                  Elad Yom{-}Tov},
  title        = {{PCL:} Prompt-based Continual Learning for User Modeling in Recommender
                  Systems},
  booktitle    = {Companion Proceedings of the {ACM} on Web Conference 2025, {WWW} 2025,
                  Sydney, NSW, Australia, 28 April 2025 - 2 May 2025},
  pages        = {1475--1479},
  publisher    = {{ACM}},
  year         = {2025},
  url          = {https://doi.org/10.1145/3701716.3715589},
  doi          = {10.1145/3701716.3715589},
  timestamp    = {Sun, 02 Nov 2025 21:27:17 +0100},
  biburl       = {https://dblp.org/rec/conf/www/YangYGXZCSLG25.bib},
  bibsource    = {dblp computer science bibliography, https://dblp.org}
}

@inproceedings{cfag,
  author       = {Mingdai Yang and
                  Zhiwei Liu and
                  Liangwei Yang and
                  Xiaolong Liu and
                  Chen Wang and
                  Hao Peng and
                  Philip S. Yu},
  title        = {Ranking-based Group Identification via Factorized Attention on Social
                  Tripartite Graph},
  booktitle    = {Proceedings of the Sixteenth {ACM} International Conference on Web
                  Search and Data Mining},
  pages        = {769--777},
  publisher    = {{ACM}},
  year         = {2023},
}

@inproceedings{gtgs,
  author       = {Mingdai Yang and
                  Zhiwei Liu and
                  Liangwei Yang and
                  Xiaolong Liu and
                  Chen Wang and
                  Hao Peng and
                  Philip S. Yu},
   title        = {Group Identification via Transitional Hypergraph Convolution with
                  Cross-view Self-supervised Learning},
  booktitle    = {Proceedings of the 32nd {ACM} International Conference on Information
                  and Knowledge Management},
  pages        = {2969--2979},
  publisher    = {{ACM}},
  year         = {2023}
}

@inproceedings{uprth,
  author       = {Mingdai Yang and
                  Zhiwei Liu and
                  Liangwei Yang and
                  Xiaolong Liu and
                  Chen Wang and
                  Hao Peng and
                  Philip S. Yu},
  title        = {Unified Pretraining for Recommendation via Task Hypergraphs},
  booktitle    = {Proceedings of the 17th {ACM} International Conference on Web Search
                  and Data Mining},
  pages        = {891--900},
  publisher    = {{ACM}},
  year         = {2024},
}

@inproceedings{Loveland25-CF,
author = {Loveland, Donald and Ju, Mingxuan and Zhao, Tong and Shah, Neil and Koutra, Danai},
title = {On the Role of Weight Decay in Collaborative Filtering: A Popularity Perspective},
year = {2025},
isbn = {9798400714542},
publisher = {Association for Computing Machinery},
booktitle = {Proceedings of the 31st ACM SIGKDD Conference on Knowledge Discovery and Data Mining V.2},
address = {New York, NY, USA},
pages = {1975–1986},
numpages = {12},
series = {KDD '25}
}

@misc{instacart,
  author       = {Kaggle},
  title        = {Kaggle Datasets by Yasser H},
  year         = {2022},
  url          = {https://www.kaggle.com/yasserh/datasets},
  note         = {Accessed: 2026-01-25}
}

@inproceedings{bonab2021crossmarket,
	author = {Bonab, Hamed and Aliannejadi, Mohammad and Vardasbi, Ali and Kanoulas, Evangelos and Allan, James},
	booktitle = {Proceedings of the 30th ACM International Conference on Information \& Knowledge Management},
	publisher = {ACM},
	title = {Cross-Market Product Recommendation},
    url       = {https://xmrec.github.io/},
	year = {2021}}

@inproceedings{Nerrise25-substitution,
author = {Nerrise, Favour and Huang, Edward W and Ji, Xiaonan and Subbian, Karthik and Koutra, Danai},
title = {GraFS: An Integrated GNN-LLM Approach for Inferring Best Functional Substitute Products},
year = {2025},
isbn = {9798400720406},
publisher = {Association for Computing Machinery},
booktitle = {Proceedings of the 34th ACM International Conference on Information and Knowledge Management},
pages = {5047–5051},
numpages = {5},
location = {Seoul, Republic of Korea},
series = {CIKM '25}
}

@InProceedings{Zhu_2025_CVPR,
    author    = {Zhu, Jing and Zhou, Yuhang and Qian, Shengyi and He, Zhongmou and Zhao, Tong and Shah, Neil and Koutra, Danai},
    title     = {Mosaic of Modalities: A Comprehensive Benchmark for Multimodal Graph Learning},
    booktitle = {Proceedings of the Computer Vision and Pattern Recognition Conference (CVPR)},
    month     = {June},
    year      = {2025},
    pages     = {14215-14224}
}

@inproceedings{Zhongmou25-Linkgpt,
author = {He, Zhongmou and Zhu, Jing and Qian, Shengyi and Chai, Joyce and Koutra, Danai},
title = {LinkGPT: Leveraging Large Language Models for Enhanced Link Prediction in Text-Attributed Graphs},
year = {2025},
isbn = {9798400720406},
publisher = {Association for Computing Machinery},
address = {New York, NY, USA},
booktitle = {Proceedings of the 34th ACM International Conference on Information and Knowledge Management},
pages = {843–853},
numpages = {11},
location = {Seoul, Republic of Korea},
series = {CIKM '25}
}

% \clearpage 
\appendix
\section{Appendix}
\subsection{Implementation Details}
Since most item titles in the three grocery datasets used in experiments are self-explainable, we use titles as textual descriptions of items and remove all the items without titles. For each dataset, users' interactions are chronologically organized based on timestamps to construct user behavior sequences. Since our framework delegates full-ranking sequential recommendation tasks to existing models, we follow the same data preprocessing strategy in previous works~\cite{LightGCN, sasrec, simplex} to remove cold-start users and items.
The recommendation tools are pretrained on the first $(n-k)$ items of each user’s interaction sequence $\mathbf{s}_{[:-k]}$, following the standard training protocol in recommendation where models learn from early interactions and are evaluated on their ability to predict future behaviors. This setup ensures consistency across tools and enables fair performance comparison between our framework and other recommendation baselines.
A grid search is applied for searching learning rates $\alpha, \beta$ and $\gamma$ in \{1e-3, 5e-3, 1e-2, 5e-2, 1e-1\}. 
To mitigate overfitting, each learning rate is multiplied by a decay factor of 0.8 after every optimization epoch. The number of historical items $c$ used for substitute and complement generation is set to 10. Following the training scheme in sequential recommendation, the number of target items $k$ is set to 1 during optimization. Since the evaluation metrics consider at most top 20 items, the number of reranking candidates $k'$ is fixed at 20. All experiments are conducted using the Phi-4 language model, deployed locally with vLLM on eight Tesla V100 GPUs, each with 32 GB of VRAM. The temperature is set to 0 to ensure deterministic outputs.
Aside from the complexity of the recommendation tools described in Sec.~\ref{sec:rectools}, the quantized version of a single Phi-4 model requires less than 12 GB of VRAM. Following prior LLM-agent works~\cite{agentcf,recagent,interecagent}, we train the recommendation tools on the entire training set, and randomly sample 160 agents for optimization on each dataset.

\begin{figure}[b!]
    \begin{subfigure}{0.163\textwidth}
    \includegraphics[width=\textwidth]{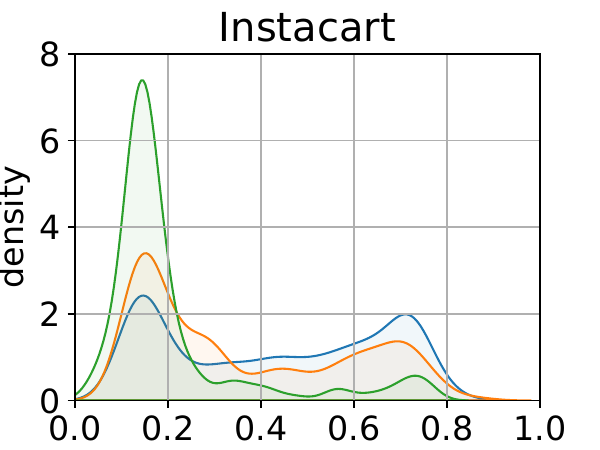}
    \end{subfigure}
    \hspace{-3mm}
    % \hfill
    \begin{subfigure}{0.163\textwidth}
    \includegraphics[width=\textwidth]{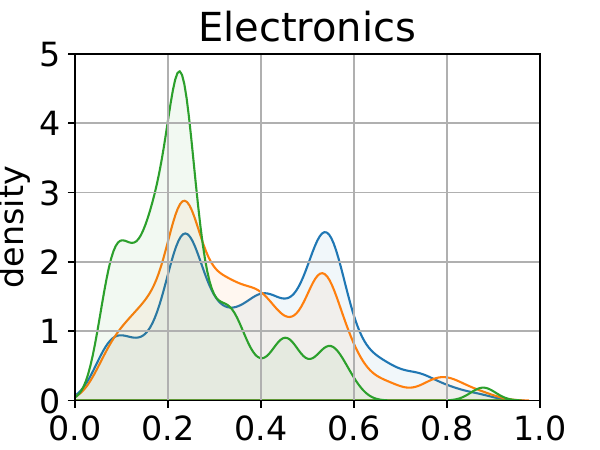}
    \end{subfigure}
    \hspace{-3mm}
    % \hfill
    \begin{subfigure}{0.163\textwidth}
    \includegraphics[width=\textwidth]{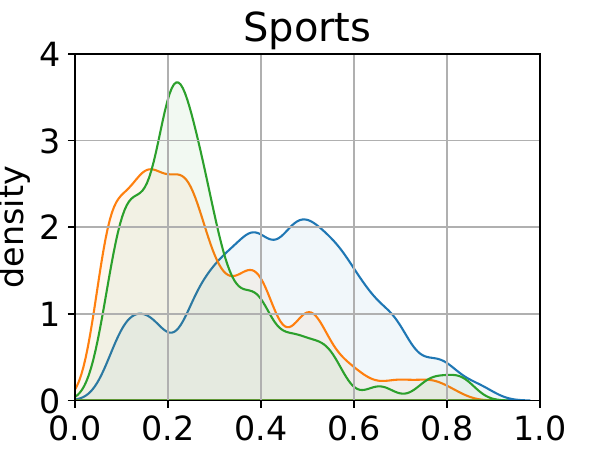}
    \end{subfigure}
    \begin{subfigure}{0.35\textwidth}
    \centering
   \vspace{-2mm}
    \includegraphics[width=\textwidth]{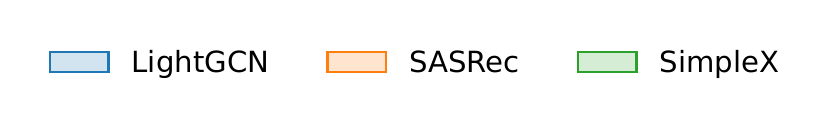}
    \end{subfigure}
    \vspace{-0.5cm}
        \caption{The KDE plot visualizing the distributions of RecTool memory in all user agents after optimization.}
  \label{fig:tool_weights}
\end{figure}

\subsection{Cost Analysis}
The total LLM API calls in \modelname consist of three parts: generation, agent optimization, and reranking. 
Generation (Eq.~\ref{eq:gen_prof}, Eq.~\ref{eq:gen_sub} and Eq.~\ref{eq:gen_com})
and reranking (Eq.~\ref{eq:gen_rerank}, Eq.~\ref{eq:rerank_sub}, and Eq.~\ref{eq:rerank_com}) contain 6 calls per user. 
Each epoch in agent optimization contains 3 calls (Eq.~\ref{eq:tool_cmpr}, Eq.~\ref{eq:scdcm} and Eq.~\ref{eq:intentdcm}) per user.
Hence, the total number of calls is $6n + 3nt$ where $n$ and $t$ are the numbers of users and epochs, respectively.
With all the agent memory updated, \modelname takes less than 1 second to generate the final ranking list for each user.
We acknowledge that LLM-based agents generally have higher inference latency than embedding-based recommendation models, but they are still practical for offline ranking or secondary-stage recommendation pipelines.

\subsection{Agent Memory Visualization}
We visualize the RecTool memories after agent optimization in Fig.~\ref{fig:tool_weights}. Kernel density estimation (KDE) plots are used to show the weight distributions of different tools for all users in the three datasets. The x-axis represents the value of the tool weight, while the y-axis reflects the density of that value. The results show that SimpleX receives lower weights than LightGCN and SASRec on Instacart and Electronics, suggesting that its low-rank factorization captures fewer semantically aligned items, whereas the latter models leverage high-order collaborative signals and temporal ordering to produce more relevant recommendations. This pattern aligns with the low VDCG of SimpleX on these two datasets in Table~\ref{tab:vdcg}.

\subsection{Details of Prompts}\label{app:prompts}
The detailed contents of prompts used in \modelname are shown from Table~\ref{app:tab:gen_prof} to Table~\ref{app:tab:dualrerank}. 
%{\color{red} ** Do you want to add the prompts for VDCG evaluation too? we have room}
The prompts for VDCG evaluation is shown in Table~\ref{app:tab:vdcg}.

\begin{promptbox}
\small
\begin{tabularx}{\linewidth}{X}
\toprule
\textbf{[Instruction]}\\
Summarize the user's preference based on the historical items this user purchased under \emph{Electronics} on \emph{Amazon}.\\[0.4em]
\textbf{[Historical Items]}\\
\{item\_descriptions\}\\
\bottomrule
\end{tabularx}
\end{promptbox}
\captionof{table}{Prompts for $\text{LLM}_\text{prof}(\cdot)$ in Eq.~\ref{eq:gen_prof} to summarize users' preferences.}
\label{app:tab:gen_prof}

% \medskip

\noindent
\begin{promptbox}
\small
\begin{tabularx}{\linewidth}{X}
\toprule
\textbf{[Instruction]}\\
According to the historical items purchased by a user, generate \emph{20} \emph{substitutes} of these items under \emph{Electronics} on \emph{Amazon}.\\[0.4em]
\textbf{[Historical Items]}\\
\{item\_descriptions\}\\[0.4em]
The output must be one list of item titles in length of \emph{20}, separated by lines.\\
\bottomrule
\end{tabularx}
\end{promptbox}
\captionof{table}{Prompts for $\text{LLM}_\text{gen}^{sub}(\cdot)$ in Eq.~\ref{eq:gen_sub} to generate substitutes. Similar prompts are used for $\text{LLM}_\text{gen}^{com}(\cdot)$ in Eq.~\ref{eq:gen_com} to generate complements.}
\label{app:tab:gen_subs}
% \medskip

\noindent
\begin{promptbox}
\small
\begin{tabularx}{\linewidth}{X}
\toprule
\textbf{[Instruction]}\\
Under \emph{Electronics} on \emph{Amazon}, according to the descriptions of items in three groups A, B and C, evaluate which group the target item is most relevant to.\\[0.4em]
\textbf{[Group A]}\\
\{item\_descriptions\}\\[0.4em]
\textbf{[Group B]}\\
\{item\_descriptions\}\\[0.4em]
\textbf{[Group C]}\\
\{item\_descriptions\}\\[0.4em]
\textbf{[Target Item]}\\
\{target\_item\_description\}\\[0.4em]
The output must be one single character in \{A, B, C\} denoting the most relevant group.\\
\bottomrule
\end{tabularx}
\end{promptbox}
\captionof{table}{Prompts for $\text{LLM}_\text{cpr}(\cdot)$ in Eq.~\ref{eq:tool_cmpr} to select the tool most aligned with the user's ground-truth interest.}
\label{app:tab:cpr}
\medskip

\noindent
\begin{promptbox}
\small
\begin{tabularx}{\linewidth}{X}
\toprule
\textbf{[Instruction]}\\
Given the two groups of items under \emph{Electronics} on \emph{Amazon}, evaluate which group is more relevant to the target item.\\[0.4em]
\textbf{[Group 1]}\\
\{item\_descriptions\}\\[0.4em]
\textbf{[Group 2]}\\
\{item\_descriptions\}\\[0.4em]
\textbf{[Target Item]}\\
\{target\_item\_description\}\\[0.4em]
The output must be one single number in \{1, 2\} denoting the more relevant group.\\
\bottomrule
\end{tabularx}
\end{promptbox}
\captionof{table}{Prompts for $\text{LLM}_\text{dcm}(\cdot)$ in Eq.~\ref{eq:scdcm} to identify whether the substitute or complement list better matches the user's ground-truth interest.}
\label{app:tab:dcm}
\medskip

\noindent
\begin{promptbox}
\small
\begin{tabularx}{\linewidth}{X}
\toprule
\textbf{[Instruction]}\\
According to the historical items purchased by a user under \emph{Electronics} on \emph{Amazon}, evaluate if this user exhibits clear substitute/complement patterns or not.\\[0.4em]
\textbf{[Historical Items]}\\
\{item\_descriptions\}\\[0.4em]
The output must be one single word in \{Yes, No\}.\\
\bottomrule
\end{tabularx}
\end{promptbox}
\captionof{table}{Prompts for $\text{LLM}_\text{dcm}^{reg}(\cdot)$ in Eq.~\ref{eq:reglearn} to evaluate if user preferences exhibit clear substitute or complement patterns.}
\label{app:tab:dcm_reg}
\medskip

\noindent
\begin{promptbox}
\small
\begin{tabularx}{\linewidth}{X}
\toprule
\textbf{[Instruction]}\\
According to the user profile, rank top-\emph{20} items this user may prefer from the candidate item list, from higher to lower probability.\\[0.4em]
\textbf{[User Profile]}\\
\{user\_profile\}\\[0.4em]
\textbf{[Candidate Item List in format of (ID, description)]}\\
\{(item\_IDs, item\_descriptions)\}\\[0.4em]
The output must be a list of candidate item IDs with length of \emph{20}, with items separated by lines.\\
\bottomrule
\end{tabularx}
\end{promptbox}
\captionof{table}{Prompts for $\text{LLM}_\text{rank}^{reg}(\cdot)$ in Eq.~\ref{eq:gen_rerank} to rerank candidate items based on their semantic similarity with the users’ general preferences.}
\label{app:tab:rank_reg}
\medskip

\noindent
\begin{promptbox}
\small
\begin{tabularx}{\linewidth}{X}
\toprule
\textbf{[Instruction]}\\
Rank top-\emph{20} items from the candidate item list based on their similarity to the target item list, from higher to lower similarity.\\[0.4em]
\textbf{[Target Item List ordered by priority]}\\
\{item\_descriptions\}\\[0.4em]
\textbf{[Candidate Item List in format of (ID, description)]}\\
\{(item\_IDs, item\_descriptions)\}\\[0.4em]
The output must be a list of candidate item IDs with length of \emph{20}, with items separated by lines.\\
\bottomrule
\end{tabularx}
\end{promptbox}
\captionof{table}{Prompts for $\text{LLM}_\text{rank}^{sub}(\cdot)$ and $\text{LLM}_\text{rank}^{com}(\cdot)$ in Sec.~\ref{sec:dualrerank} to rerank candidate items based on their semantic similarity with the potential substitutes or complements.}
\label{app:tab:dualrerank}

\noindent
\begin{promptbox}
\small
\begin{tabularx}{\linewidth}{X}
\toprule
\textbf{[Instruction]}\\
Given the same category, rate how well each item from the recommended list matches the target item based on relevance, usefulness, and user interest.\\[0.4em]
\textbf{[Category]}\\
\{item\_category\}\\[0.4em]
\textbf{[Target Item]}\\
\{item\_description\}\\[0.4em]
\textbf{[Recommended List]}\\
\{item\_descriptions\}\\[0.4em]
The output must be a list of integers. Each score must be between 0 (very unrelated) and 9 (exact match).\\
\bottomrule
\end{tabularx}
\end{promptbox}
\captionof{table}{Prompts for the LLM to rate how well each item in the recommended list aligns with the ground-truth item for VDCG evaluation in Sec.~\ref{sec:vdcg}.}
\label{app:tab:vdcg}

% \begingroup
% \let\origtable\table
% \let\endorigtable\endtable
% \renewenvironment{table}[1][]{\origtable[]}{\endorigtable}

% \endgroup

% \begin{table}
% \centering
% \begin{promptbox}
% \begin{tabularx}{\linewidth}{|X|}
% \hline
% [Instruction] \\
% Given the same category, rate how well each item from the recommended list matches the target item based on relevance, usefulness, and user interest.\\
% \vspace{0.4em}
% [Category]\\
% \{item\_category\}\\
% \vspace{0.4em}
% [Target Item]\\
% \{item\_description\}\\
% \vspace{0.4em}
% [Recommended List]\\
% \{item\_descriptions\}\\
% \vspace{0.4em}
% The output must be a list of integers. Each score must be between 1 (very unrelated) and 10 (exact match).\\
% \hline
% \end{tabularx}
% \end{promptbox}
% \caption{Prompts for the LLM to rate how well each item in the recommended list aligns with the ground-truth item in Sec.~\ref{sec:vdcg}.}\label{app:tab:vdcg}
% \end{table}

\subsection{Additional Related Work}~\label{app:related_works}

In this appendix, we discuss additional works related to ours, including research on LLM-based recommendation systems and methods for identifying substitute and complementary products.

\subsubsection{LLM for Recommendation}
The capabilities of traditional embedding-based recommender systems remain limited when relying solely on historical interaction data. To address this limitation, prior works have incorporated LLMs either as text encoders to enrich recommendation embeddings with textual information~\cite{llmrec, rlmrec,pcl,ihp,Zhu_2025_CVPR,Nerrise25-substitution} or as standalone recommenders by reformulating recommendation tasks as language modeling problems~\cite{p5, m6rec,Zhongmou25-Linkgpt}.
While these approaches leverage the world knowledge of LLMs and show advantages in cold-start scenarios~\cite{HuangBYJ0SWKY25, llmrank}, their effectiveness is constrained by the fixed dimensionality of feature vectors and the fundamental mismatch between language and behavior modeling, which hinders their ability to fully exploit interaction signals. 
We compare these pioneer works with LLM-based recommendation agents~\cite{agentcf,agent4rec,iagent,interecagent,recagent,recmind} in Table~\ref{tab:salesman}.

To tightly couple the pretrained knowledge in LLMs and ID-based behavior patterns in recommender systems, recent works introduce user and item tokens into the LLM vocab space, and align these tokens with the language space~\cite{LiaoL0WYW024, ZhuWGHL24, YangLYLW0Y25}. However, compared to inference-only LLM agents for recommendation, the effectiveness of this knowledge alignment with LLMs is sensitive to the amount of tunable parameters and available data.
Besides, the concept of Retrieval-Augmented Generation (RAG) has been adapted for recommendation, where candidate interactions are first retrieved without LLMs and then reasoned by LLMs~\cite{kgrag, coral}. Retrieval alleviates the reasoning load for LLMs by narrowing the candidate space, but the overall performance is influenced by the inherent biases of the specific deployed retrieval method. In contrast, LLM agents usually leverage multiple tools to enhance the performance on the target task, like \modelname equipped with full-ranking tools for recommendation.

\subsubsection{Substitution and Complementarity}
Although substitution and complementarity as item-item relations play a crucial role in understanding user intent~\cite{subcom22, subcom23,Nerrise25-substitution}, 
% Substitutes help identify alternatives when a preferred item is unavailable or when the user seeks variety, while complements reveal co-purchase patterns that often reflect planned consumption behaviors.
they are unavailable in most public datasets~\cite{lacklabel19, lacklabel21}. 
As a result, previous works~\cite{sclabel20chen, sclabel20liu, sclabel20xu} constructed these labels from co-view and co-purchase logs of users due to the inefficiency of rigorous label annotation by human annotator. 
However, many item-pairs labeled based on these behavior logs are not functionally substitution or complementary~\cite{notrealsc22tong, notrealsc23ye, notrealsc24kai}. To address this gap, LLMs have been applied for labeling substitution and complementary relations in previous works~\cite{llmlabel23papso, llmlabel25Hosseini}, indicating that their extensive world knowledge make LLMs particularly well-suited for reasoning over these implicit item-item relationships in e-commerce datasets. In \modelname, these relationships are leveraged during intent discrimination and relational reranking to enhance recommendation relevance.

\end{document}